\documentclass[prb,amsmath,amssymb,twocolumn,superscriptaddress,showpacs]{revtex4-1}
\usepackage{bm}
\usepackage{amssymb}
\usepackage{colordvi}
\usepackage{graphicx}
\usepackage{color}
\usepackage{hyperref}
\newcommand{\up}{\uparrow}
\newcommand{\down}{\downarrow}

\newcommand{\be}{\begin{equation}}
\newcommand{\ee}{\end{equation}}
\newcommand{\bea}{\begin{eqnarray}}
\newcommand{\eea}{\end{eqnarray}}

\newcommand{\vep}{\varepsilon}

\newcommand{\ome}{\omega}

\def\Re {\mbox{Re}}
\def\Im {\mbox{Im}}

\def\nn{\nonumber}
\def\pp{\parallel}

\def\grad {\mbox{\boldmath$\nabla$\unboldmath}}

\begin{document}

\title{Three-dimensional photonic Dirac points stabilized
  by point group symmetry}

\author{HaiXiao Wang}
\author{Lin Xu}
\author{HuanYang Chen}\email{chy@suda.edu.cn}
\author{Jian-Hua Jiang}\email{jianhuajiang@suda.edu.cn}
\affiliation{College of Physics, Optoelectronics and Energy, \&
  Collaborative Innovation Center of Suzhou Nano Science and
  Technology, Soochow University, 1 Shizi Street, Suzhou 215006,
  China}

\date{\today}

\begin{abstract}
We discover a pair of stable 3D Dirac points, 3D photonic analog of
graphene, in all-dielectric photonic crystals using structures
commensurate with nano-fabrication for visible-frequency photonic
applications. The Dirac points carry nontrivial $Z_2$ topology and
emerge for a large range of material parameters in hollow cylinder
hexagonal photonic crystals. From Kramers theorem and group 
theory, we find that only the $C_6$ symmetry lead to
point group symmetry stabilized Dirac points in 3D all-dielectric
photonic crystals. {The Dirac points are characterized using ${\vec
  k}\cdot{\vec P}$ theory for photonic bands in combination with
symmetry analysis. Breaking inversion symmetry splites the Dirac
points into Weyl points. The physical properties and experimental
consequences of Dirac points are also studied. The Dirac points are
found to be robust against parameter tuning and weak disorders.}
\end{abstract}


\maketitle

\section{Introduction}
Theoretical predictions and experimental
discoveries of quantum (spin) Hall effect and Weyl points (WPs) in photonic
crystals (PCs) stimulated the study of topological properties of photonic
systems as well as their
applications.\cite{haldane1,taylor1,mit1,mit2,austine,ling1,ling-exp,rev,nori}
For example, WPs induce unique surface states with chiral
isofrequency contour\cite{ling1,ct-exp,szhang} and strongly modify
photon polarization when light is reflected by the surface of a medium with
WPs.\cite{szhang} 

Unlike WPs that exist only in systems with broken time-reversal
symmetry (TRS) or inversion symmetry (IS), 3D Dirac points (DPs) exist
in systems with concurrent TRS and IS.\cite{zhong1,zhong2} 3D
topological DPs, have attracted marked research interest
recently\cite{zhong1,zhong2,chen,furusaki,nagaosa,ong,cfang} as they
exhibit a wide variety of anomalous effects (even not shared with WPs)
such as linear quantum magnetoresistance,\cite{ong} quantum spin Hall
effect,\cite{zhong1} and strong diamagnetism.\cite{zhong1} DPs are
important also because they are parent states of various topological
states. For example, gap formation in DPs can lead to topological
insulators.\cite{zhong1,nagaosa,kane,hu,ling2}

A DP consists of a pair of WPs with opposite chirality (i.e., a DP is
a four-fold degenerate point around which the effective Hamiltonian
resembles that of the famous Dirac equation), which is
usually unstable as the two WPs can annihilate each other and form a
gap. It was found only recently that a pair of DPs become stable with
certain point group symmetry.\cite{zhong1,nagaosa} In electronic
systems a DP consists of two-fold spin degeneracy and two-fold accidental
orbital degeneracy. 

In photonic systems, however, due to the
fundamental distinction of Kramers theorem for fermions and bosons
(i.e., ${\cal T}^2=-1$ for fermions with ${\cal T}$ being
time-reversal operation induces double degeneracy, whereas ${\cal
  T}^2=1$ for bosons lack such degeneracy), there is no spin-
(polarization-) degeneracy. {It was pointed out in
Ref.~\onlinecite{lumh} that the electromagnetic duality symmetry of
the Maxwell equation can be exploited to generate the two-fold
polarization degeneracy since the duality transformation {${\cal D}:
({\vec E}, {\vec H})\to (\vec{H},-{\vec E})$; ${\vep}\to \mu$}
satisfies ${\cal D}^2=-1$. However, such a scheme requires
bianisotropic medium with $\vep=\mu$ which is 
difficult to achieve particularly for optical frequencies.
In all-dielectric 3D PCs which explicitly break
electromagnetic duality symmetry\cite{lumh} (since $\vep\ne\mu$),
spin- (polarization-) degeneracy of photon without 
fine-tuning is unlikely.\cite{book} For instance, there are only
two-fold degeneracy for the $K$ point (and another two-fold degeneracy
for the $K^\prime$ point) for photonic graphene with both TRS and
IS. In comparison there are four-fold degeneracy for the $K$ point
(and another four-fold degeneracy for the $K^\prime$ point) for
electronic graphene.} The distinction is due to the absence of Kramers
double degeneracy (i.e., spin-degeneracy) in photonic systems. A 3D
photonic DP hence requires four-fold orbital degeneracy. It is unclear
whether such accidental degeneracy is stable against perturbations.
Moreover, designing topological states in all-dielectric photonic
crystals is much more difficult than in electronic band materials due
to the nature of photonic bands: they are mainly formed by multiple
coherent scattering\cite{sajeev,eli} rather than local atomic orbits.

In this work, we discover a pair of DPs with equal frequency (i.e., in
total eight-fold degeneracy at such frequency) in all-dielectric PCs 
whose structure is commensurate with nano-fabrications for
visible-frequency applications. Such {\em paired} DPs, as genuine
analog of 3D photonic graphene, have never been found or studied in
the literature of photonics, although the study of Weyl points and
other topological phenomena in photonics has attracted a lot of
attention. Remarkably, the DPs carry nontrivial $Z_2$ topology and are
stable in a large material-parameter region. We show that $C_6$ is the
only point group symmetry that can stabilize the {\em paired} DPs in
3D PCs. Here the {\em paired} DPs are distinct from {\em unpaired} DPs
locating at high-symmetry points of the Brillouin zone due to
fine-tuned parameters\cite{2dmat} or nonsymmorphic
symmetries\cite{ling2} or Dirac-like points with 
three-fold degeneracy.\cite{zim3d} As 3D photonic analog of graphene,
the DPs can be exploited for various graphene applications such as
Klein tunneling\cite{klein} and suppressed back-scattering which will
be useful for robust signal transmission. Since genuine photonic
devices are 3D systems,\cite{pnas} 3D photonic DPs are superior for
applications than their 2D counterparts.\cite{2dg1,2dg2,2dg3,zim}
{The paired DPs can also be exploited as parent states to design other
photonic topological states, such as 3D topological insulators of
light. We interpret the DPs as accidental but unavoidable degeneracy
points between the $p$- and $d$-like photonic bands, establishing
an ``atomic orbits'' picture using Mie resonances for the design of 
topological states in photonics. Our study provides effective
methodology beside the discovery of 3D $Z_2$ DPs in photonic systems.}

\begin{figure}
\begin{center}
\includegraphics[width=8.4cm]{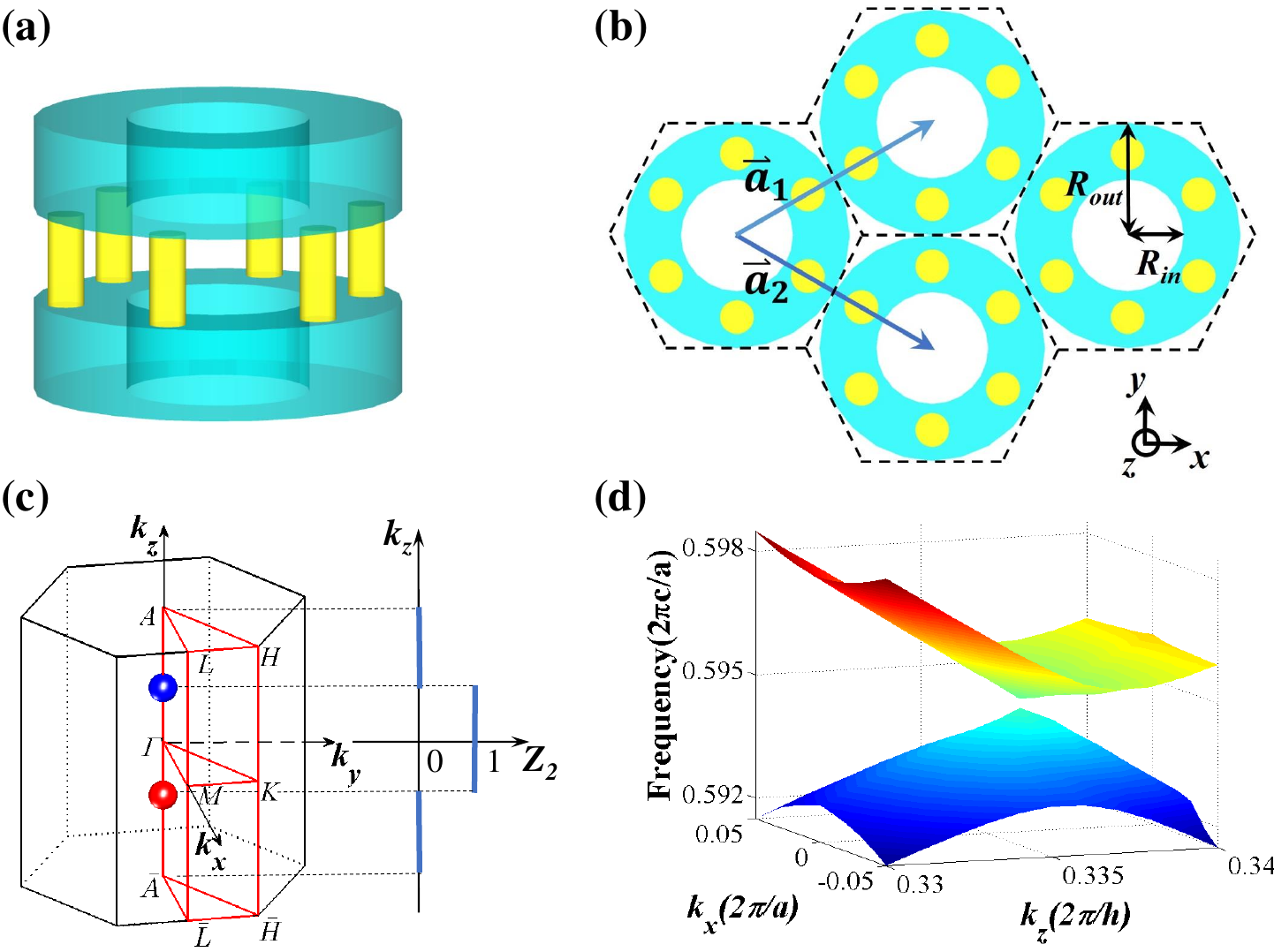}
\caption{ (Color online) (a) Structure in real-space unit cell of 
  hexagonal PCs. Hollow cylinders and micropillars are
  made of the same material with isotropic permittivity. (b) Top-down
  view of hexagonal PC. ${\vec a}_1$ and ${\vec a}_2$ are the two
  lattice vectors. $a\equiv |\vec{a}_1|=|\vec{a}_2|$ is the lattice
  constant in the $x$-$y$ plane. The height of each unit cell is
  $h=0.6a$. The diameter of each micropillar is $0.1a$ and its height
  is $0.2a$. The outer and inner radii of the hollow cylinder are $R_{out}$ and
  $R_{in}$, respectively, whereas its height is $0.4a$. (c) First Brillouin
  zone with a pair of DPs along the $\Gamma$-$A$ line as kinks of the
  $Z_2$ number vs. $k_z$. A DP with $n_D=1$ (-1) is
  labeled as a red (blue) point. (d) Dispersion close to a DP.}
\end{center}
\end{figure}

\section{Hexagonal photonic crystals with $C_6$ symmetry}
The hexagonal PC consists of hollow cylinders (with inner and outer
radii $R_{in}$ and $R_{out}$, respectively) connected by 
micropillars [Figs.~1(a)--1(b)]. The height of each unit cell is
$h=0.6a$ with $a$ being the lattice constant in the $x$-$y$ plane. The
micropillars are of the same height $0.2a$ and diameter $0.1a$. There
are six micropillars in each unit cell with arrangement preserving
$C_6$ symmetry. The height of each hollow cylinder is $0.4a$. The
Brillouin zone and high-symmetry points are depicted in Fig.~1(c). A
pair of DPs emerge at $(0,0,\pm K_z)$. The DPs carry nontrivial $Z_2$
topology: They are kinks of the $Z_2$ number vs. $k_z$ [Fig.~1(c)].
A topological number $n_D=\pm 1$ is associated with each kink, which
is defined as the change of $Z_2$ number across the kink (with
increasing $k_z$). {The DPs are actually monopoles of the $SU(2)$
Berry-flux. They are quite different from the Weyl points in photonic
crystals\cite{ling1,ct-exp,szhang} which are monopoles of the $U(1)$
Berry-flux and have a $Z$ topological charge. Our work is the first
proposal of the $Z_2$ topological Dirac points in photonic crystals. }

Topological DPs appear in pairs: each pair consists of two
DPs of opposite $n_D$ ($n_D$ can only be $\pm 1$ since $Z_2$ can only
be 0 or 1), wavevector but the same frequency
[Fig.~1(c)]. The linear dispersion around a DP is shown in
Fig.~1(d). A survey of photonic bands in the first Brillouin zone is
shown in Fig.~2(a). Using permittivity $\vep=12$ for 
dielectric materials and $R_{out}=0.5a$ 
and $R_{in}=0.4a$, our calculation [using MIT photonic 
bands (MPB)] gives $K_z=0.338\frac{2\pi}{h}$. The frequency of the DPs
is about $0.6\frac{2\pi c}{a}$ (above the light-line).

Here DPs originate from accidental crossing of the $p$ and $d$ bands
[Fig.~2(a)]. Such crossing is unavoidable if the order of the $p$ and
$d$ bands in frequency are switched at the $\Gamma$ and $A$
points. Usually the $p$ bands have lower frequency than the $d$
bands. The inversion of the $p$ and $d$ bands results in quantum spin
Hall effect ($Z_2=1$) for each $k_z$ with $|k_z|<K_z$ (see
Appendix~A).\cite{hu} The double degeneracy of the
$p$ and $d$ bands along the $\Gamma$-$A$ line is guaranteed by the
$C_6$ symmetry\cite{sakoda} [their field profiles at $\Gamma$ point
are shown in Figs.~2(b) and 2(c)]. Interestingly, the polarization of the
$p$ and $d$ bands are mostly $E_z$-like. This property is consistent
with the picture that Mie resonances of 
hollow cylinders can be regarded as photonic ``atomic orbits''
from which photonic bands are derived.\cite{soukoulis} Indeed, in
hollow cylinders the frequency of Mie resonances for TM polarization
is lower than for TE (see Appendix~B). We note that the DPs along the
$\Gamma$-$A$ line exist even when the micropillars are
removed. In this situation, however, another DP emerges at the $K$ and
$K^\prime$ point (see Appendix~C).

{Due to the complexity of photonic energy bands, up till now the design
of topological properties in photonic crystals remains accidental and
challenging particularly for three-dimensional photonic crystals. This
is due to the essential difference between electronic  and photonic
energy bands: The photonic energy bands are consequences of multiple
Bragg scattering of the vectorial electromagnetic waves as no
dielectric material can trap light\cite{sajeev,eli,book}. In contrast,
electronic band structure can mostly be understood as hybridization of
local atomic orbits. For materials with inversion symmetry, the $Z_2$
topological index can be calculated simply by counting band (parity)
inversion at high symmetry points in the Brillioun zone. Such a
simplified picture is not available in photonic crystals, creating
lots of obstacles in designing and understanding the topological
properties of photonic bands. }

{Our design is based on that the Mie resonances of hollow cylinders can
be regarded as photonic local orbits. The photonic energy bands can be
understood as derived from photon hopping between Mie resonances in
neighboring unit cells\cite{soukoulis} (except for the lowest
two photonic bands which also consist of the plane-wave component, see
Appendix~B). The hollow cylinder supports
$p$- and $d$-wave Mie resonances with double degeneracy as protected
by the $C_{6}$ and inversion symmetries\cite{sakoda,hu}. Exploiting
such double degeneracy a pseudo-time-reversal operation ${\cal
  T}_p=i\hat{\sigma}_y{\cal T}$ with ${\cal T}_p^2=-1$ can be defined
where $\hat{\sigma}_y$ is the pseudo-spin Pauli matrix\cite{hu} (the
double degenerate Hilbert space is defined as the pseudo-spin). This
degeneracy property has been exploited to construct 2D photonic
topological insulators recently\cite{hu}. In our hollow
cylinder photonic crystals the $p$- and $d$-like bands can be viewed as
derived from the $p$- and $d$-wave Mie resonances. A pair of 
accidental but unavoidable degeneracy points of the $p$- and $d$-
photonic bands are discovered, which are identified as the $Z_2$
topological DPs. The degeneracy of the $p$- and $d$- bands comes from
space group symmetry rather than the spin (polarization) degree of
freedom of photon. The topological property of those photonic bands is
the same as that of the energy bands in spinless bosonic systems.}

\begin{figure}
\begin{center}
\includegraphics[width=8.4cm]{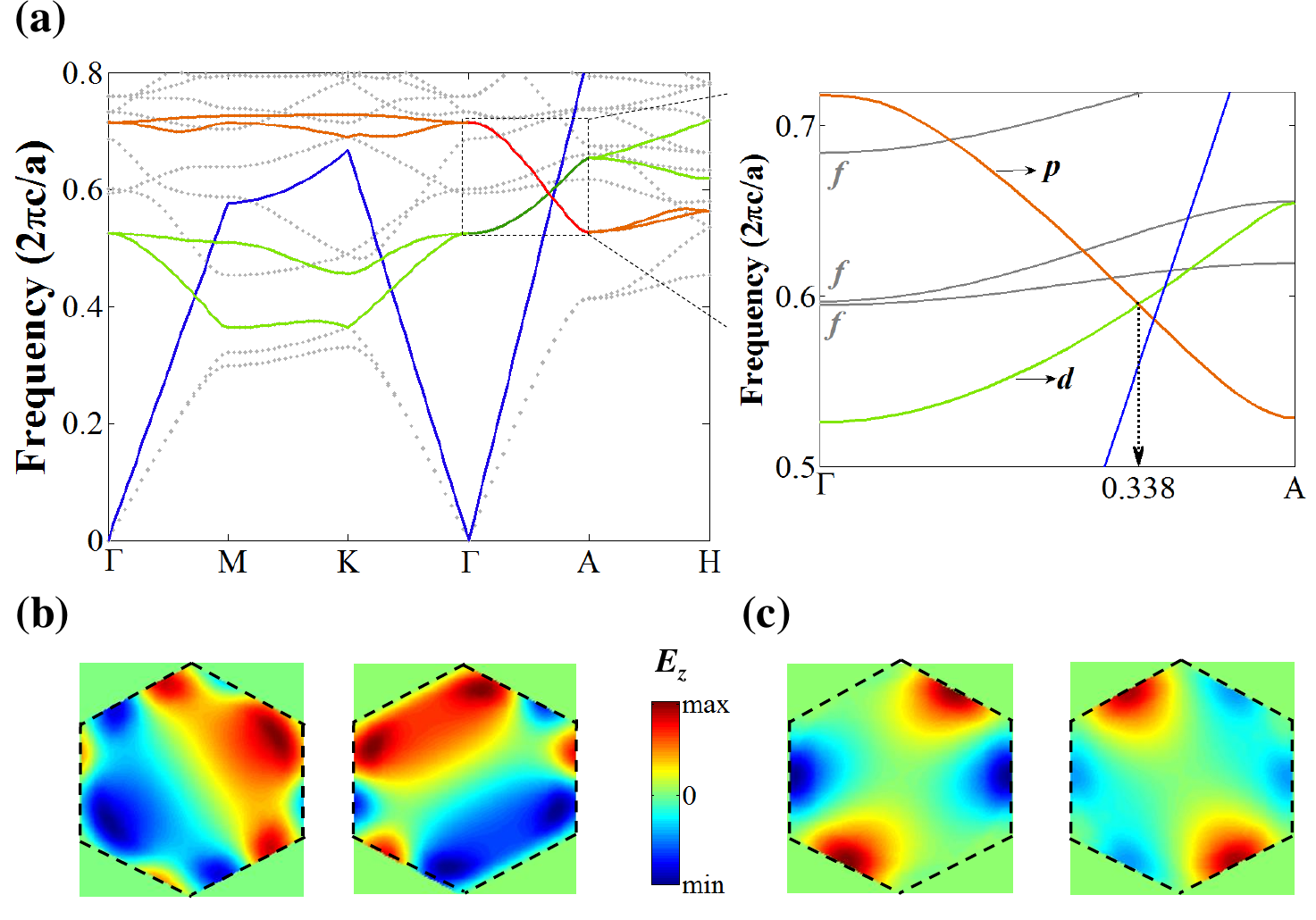}
\caption{ (Color online) (a) Photonic band structure of a hexagonal PC
  with $C_6$ and inversion symmetry (blue curve indicating light-line).
  Zoom-in: Band structure along $\Gamma$-$A$ line. The $p$-bands
  (red) cross the $d$-bands (green) at $(0,0,\pm K_z)$ with
  $K_z=0.338\frac{2\pi c}{h}$. The gray curves represent the
  $f$-bands. (b)-(c): $E_z$ field profile in the $x$-$y$ plane of (b)
  $d$-bands and (c) $p$-bands at $\Gamma$ point. A unit cell is
  depicted by the black dashed curves. Parameters: $R_{out}/a=0.5$
  and $R_{in}/a=0.4$, and permitivity $\vep=12$.} 
\end{center}
\end{figure}

\section{Photonic ${\vec k}\cdot{\vec P}$ Hamiltonian and point group
  symmetry analysis}

Near each $(0,0, k_z)$ point the doubly degenerate $d$
bands, $d_{x^2-y^2}$ and $d_{xy}$, can be reorganized into the
pseudo-spin-up $d_{+}=d_{x^2-y^2}+id_{xy}$ and spin-down
$d_{-}=d_{x^2-y^2}-id_{xy}$ states (similarly for the $p$
bands).\cite{hu} Using symmetry and ${\vec k}\cdot{\vec P}$
analysis [see Appendix~A],\cite{sakoda} the photonic Hamiltonian near
the DPs $(0,0,\alpha K_z)$ ($\alpha=\pm 1$) is constructed as
\begin{align}
&\hspace{-0.15cm} \hat{{\cal H}}_\alpha = \frac{\ome_D^2}{c^2}\hat{1} +
\frac{2\ome_D}{c^2}\{\alpha (v_0 q_z\hat{1} + v_z q_z \hat{\tau}_z) \nn\\
& \hspace{0.8cm}+
[(\Re{v_{\parallel}}+i\hat{\sigma}_z\Im{v_{\parallel}})(
k_x-i\hat{\sigma}_zk_y) \hat{\tau}_+  + {\rm H.c.}]
\}. \label{hami}
\end{align}
where $\ome_D$ is the frequency of the DP, $v_0=(v_d+v_p)/2$
and $v_z=(v_d-v_p)/2$ with $v_d>0$ and $v_p<0$ being the
group velocity of the $d$ and $p$ bands at $(0, 0, K_z)$,
respectively. $q_z=k_z-\alpha K_z$, and $|v_{\parallel}|$ is the 
group velocity in the $x$-$y$ plane. $\hat{\tau}_z$ has eigenvalue 1 (-1)
for the $d$ ($p$) band, whereas
$\hat{\tau}_+=(\hat{\tau}_x+i\hat{\tau}_y)/2$. $\hat{\sigma}_z$ is the
Pauli matrix for pseudo-spin. The photonic spectrum 
$\ome_j({\vec k})$ ($j$ labeling the band index) is 
related to the eigenvalue $\lambda_j$ of the Hamiltonian via
$c^2\lambda_j=\ome_j^2({\vec k})$. For $|k_z|<K_z$, the above
Hamiltonian resembles that of the quantum spin Hall
insulator ($Z_2=1$),\cite{bhz} otherwise the band topology is trivial
($Z_2=0$). Eq.~(\ref{hami}) represents a
class of massless Dirac Hamiltonian.

\begin{figure}
\begin{center}
\includegraphics[width=8.4cm]{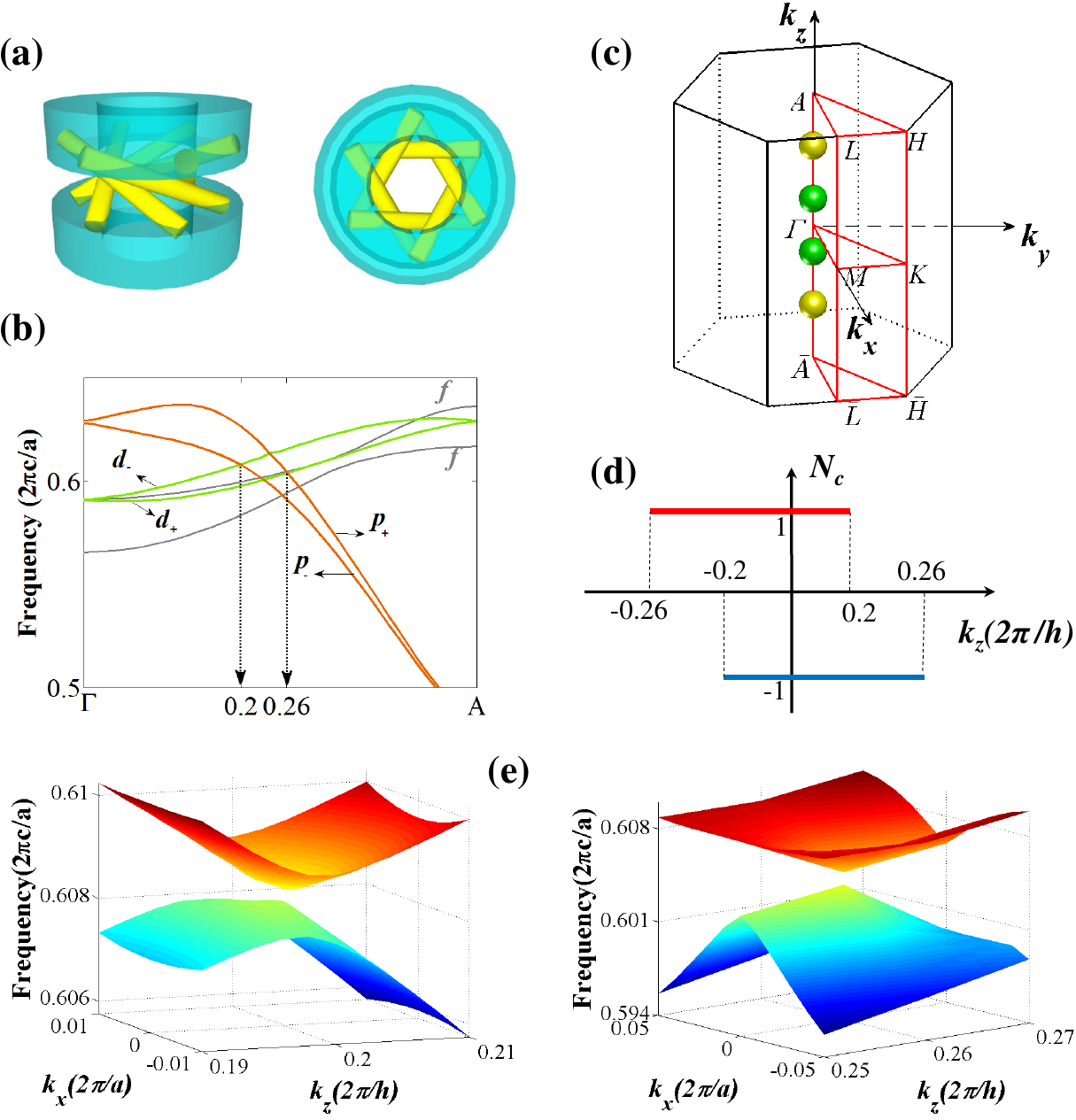}
\caption{ (Color online) IS broken hexagonal PC with $C_6$ symmetry. (a)
  Lateral (left) and top-down (right) views of the structure in a
  real-space unit cell. (b) Photonic bands along the $\Gamma$-$A$ line
  with $k_z>0$. Two WPs are found at (0, 0, $K_{z1}$) and (0,  0,
  $K_{z2}$) with $K_{z1}=0.2\frac{2\pi c}{h}$ and
  $K_{z2}=0.26\frac{2\pi c}{h}$, respectively. Other two WPs in the
  $k_z<0$ region are not shown. (c) Depicting the four WPs in
  the first Brillouin zone. Green (yellow) spheres denote WPs with
  chirality -1 (+1). (d) Chern number $N_c$ for the $p_+$-$d_+$ bands
  below the WPs (blue) and for the $p_-$-$d_-$ bands below the WPs
  (red). (e) Photonic spectrum near the two WPs. Left: for
  WP at (0, 0, $K_{z1}$); right: for WP at (0, 0, $K_{z2}$). }
\end{center}
\end{figure}

Since fine-tuned four-fold degeneracy's at particular high-symmetry
points are unstable, we consider DPs located on a symmetric line
in the Brillouin zone which has a point group symmetry (a subgroup of
the symmetry group of the PC). Such point
group symmetry can be rotation along the symmetric line or mirror with
respect to a plane containing the line. From group theory, only $C_6$
(or $C_{6v}$) group contains two doubly-degenerate representations,
which may allow four-fold accidental degeneracy.\cite{sakodabook}
Therefore, point group symmetry stabilized DPs can
only appear in hexagonal PCs with $C_6$ (or $C_{6v}$) symmetry.
There are two possible cases that fulfill such requirements: DPs on
the $\Gamma$-$A$ line, and DPs on the $K$-$H$ line (combined with
$K^\prime$-$H^\prime$ line to restore the $C_6$ symmetry).
For both cases, detailed ${\vec k}\cdot{\vec P}$ analysis
reveals that the $C_6$ point group symmetry governs the form of
Hamiltonian and guarantees $Z_2$ topology of the DPs (see
Appendix~D). Therefore, among all point group symmetry,
only $C_6$ and $C_{6v}$ can stabilize the DPs. 
This property of photonic bands is in
sharp contrast with electronic bands where several
classes of point group symmetry (including $C_3$, $C_4$, and $C_6$)
can stabilize the DPs.\cite{nagaosa} The essential difference is that
there is no two-fold spin degeneracy of photon in all-dielectric PCs,
due to its bosonic nature and due to the breakdown of duality symmetry.

An IS breaking mechanism is introduced by twisting the
micropillars and extending their heights to $0.4a$ [Fig.~3(a)]. The
degeneracy between the $p_+$ and $p_-$ bands as well as that for the
$d_+$ and $d_-$ bands are now lifted [Fig.~3(b)]. However, since the
$C_6$ symmetry is kept, the coupling between those bands remain the same
form as in (\ref{hami}). The $C_6$ symmetry guarantees that those four
bands can cross each other, since they have different $C_6$
eigenvalues. Therefore, the IS breaking mechanism only introduces the
following perturbation
\be
\delta H = \frac{2\ome_D}{\alpha c^2}(m_1 \hat{\sigma}_z + m_2 \hat{\tau}_z + m_3
\hat{\tau}_z\hat{\sigma}_z) ,
\ee
where $m_1, m_2$, and $m_3$ are small parameters. The DPs are 
split into four WPs [Fig.~3(c)] with unique topology [Fig.~3(d)]:
The Chern number of the pseudo-spin-up (-down) bands below the WP is
-1 (1) for $k_z$ between $-0.2\frac{2\pi}{h}$ ($-0.26\frac{2\pi}{h}$)
and $0.26\frac{2\pi}{h}$ ($0.2\frac{2\pi}{h}$). The total Chern number
is nonzero only for $0.2<|k_z|h/(2\pi)<0.26$. This
special distribution of Chern number is a signature of $Z_2$
topological DPs. This unique property also influences the chiral
surface states induced by the WPs as shown in the Appendix~E. {There we
show that chiral surface states can appear at the boundary between two
IS broken photonic crystal with opposite chirality. The chiral
surface states can be observed in experiments to verify our
predictions here. The $k_z$ dependent one-way edge states at such
boundaries can be exploited for non-reciprocal waveguides. Recently,
photonic edge states are exploited for protected quantum-entanglement
propagation\cite{qe1,qe2}. }

{The spectrum of the two WPs with $k_z>0$ is shown in Fig.~3(e). The
spectrum for $k_z<0$ are related with that for $k_z>0$ via the
TRS. For example, $p_+$ state with wavevector ${\vec k}$ has the same
frequency as $p_-$ state with wavevector $-{\vec k}$. The WPs
connected by TRS are of the same chirality.}

\begin{figure}
\begin{center}
\includegraphics[width=8.6cm]{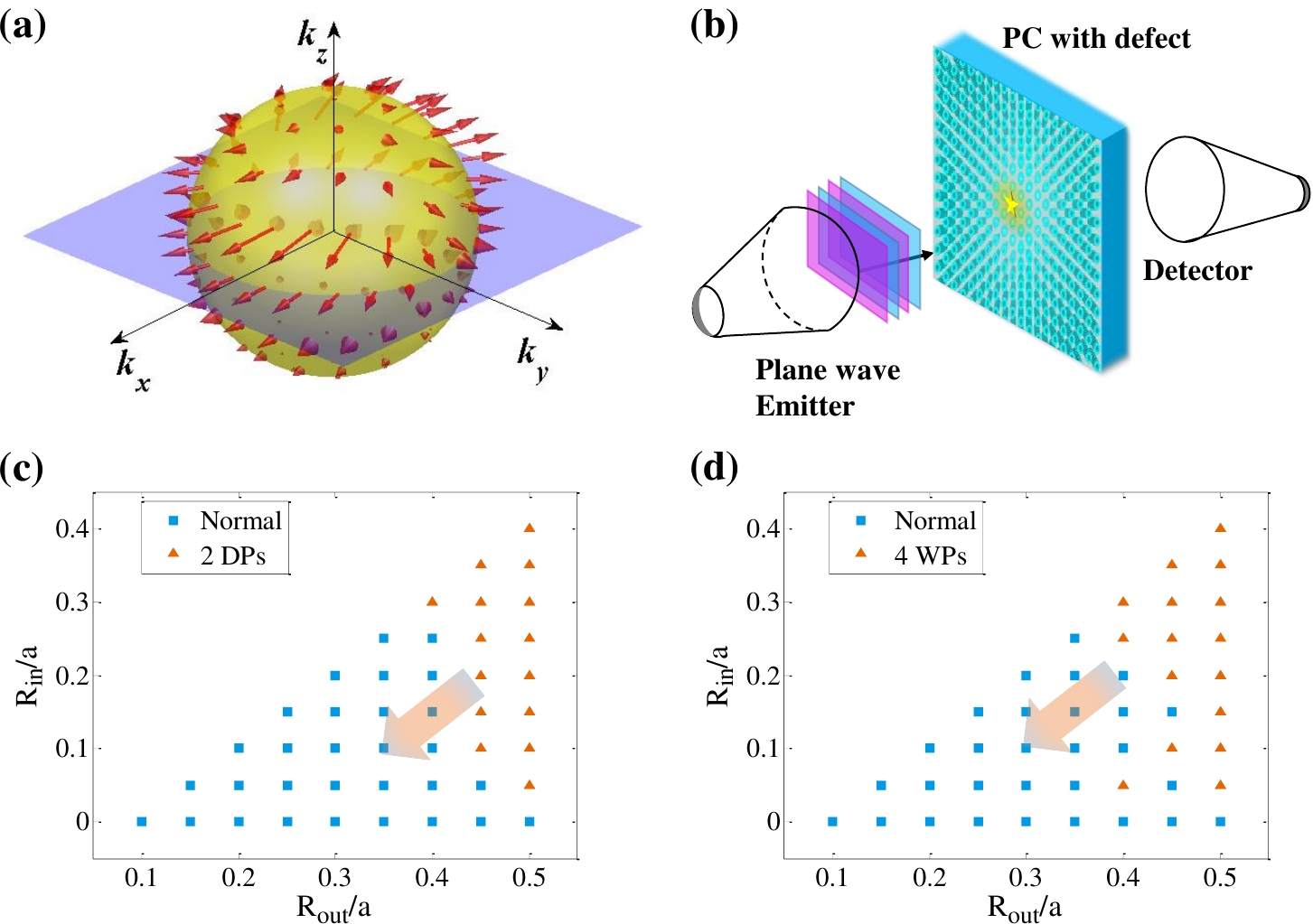}
\caption{ (Color online) (a) Angular-momentum distribution on a sphere
  surrounding the WP at $(0,0, -0.2\frac{2\pi c}{h})$ for an IS broken
  PC (same parameters as in Fig.~3). Calculated for the band below the
  WP. (b) Measurement set-up for back-scattering suppression in a PC
  with DPs. If the small defect at the center of the PC preserves
  $C_6$ symmetry, the back-scattering is suppressed; otherwise, the
  back-scattering considerably reduces transmission. (c)-(d): Phase 
  diagrams of hollow-cylinder hexagonal PCs with (c) and without
  (d) IS for various $R_{out}$ and $R_{in}$.}
\end{center}
\end{figure}

\section{Physical consequences and robustness of Dirac and Weyl points}
One salient feature of a DP is the angular-momentum(AM)--wavevector
locking:\cite{zhong1,zhong2,chen,ong} around the DP the direction of
AM is uniquely determined by the wavevector. Note that away from the
DP the doubly degenerate bands have opposite AM according to
concurrent TRS and IS. To visualize the AM-wavevector locking, one
needs to break, e.g., IS. In Fig.~4(a) we plot the distribution of AM
on a sphere surrounding the WP at (0,0,-0.2$\frac{2\pi}{h}$) for the
IS broken PC studied in Fig.~3. The correlation between the AM 
and the wavevector is clearly visible. The winding geometry of AM on
the sphere convinces the chirality of the WP. When the radius of the
sphere is small enough ($\le 0.01\frac{2\pi}{a}$) the AM distribution
does not vary with the radius. {Such $4\pi$ winding angle of AM 
is a topological property of the WP with chirality $+1$.}

AM-wavevector locking is at the heart of various novel properties of
graphene and Dirac semimetal\cite{zhong1,zhong2,chen,ong} such as
Klein tunneling,\cite{klein} pseudodiffusive
transport,\cite{2dg1,2dg3} suppressed back-scattering,\cite{ong} and
{\em Zitterbewegung} of photon.\cite{2dg2} Suppression of
back-scattering can be experimentally verified via the transmission
measurement illustrated in Fig.~4(b). A small defect embedded in the
PC may deflect the incident light depending on its symmetry. If the
defect preserves the $C_6$ symmetry, then scattering is suppressed.
In comparison, a defect with $C_3$ or $C_1$ symmetry can mix all of
the four nearly degenerate modes around the DP and
break down the protection from back-scattering.
The above phenomenon can be used as another experimental
signature (via comparing the transmission contrast) of the DPs.

{The field profiles can also be exploited to identify the topological
WPs in the IS-broken photonic crystals [see Appendix~F]. We found that
the Poynting vector exhibits spatial distributions similar to the
skyrmion configuration. The winding direction of the Poynting vector
coincides with the direction of orbital angular momentum (i.e., along
$z$ direction).}

To study the robustness of the
DPs and WPs, we calculate the phase diagram of the hexagonal PCs with
various inner and outer radii for both PCs with and without IS.
The results are shown in Figs.~4(c) and 4(d). We find that a pair of
DPs emerge for a large range of parameters, indicating robustness of
the DPs. In fact, the DPs can only be annihilated or created at the
$\Gamma$ or $A$ point, which is the physical origin of their
robustness. Since WPs are derived from the DPs, they are also robust
and emerge in a large region of material parameters. Arrows in
Figs.~4(c) and 4(d) indicate the tendency that the DPs or WPs move
toward the $\Gamma$ point when the outer or inner radius is reduced
[see Appendix~G]. {The DPs are also robust to weak disorders such as
fabrication errors, as discussed in Appendix~H. We simulated the
fabrication resolution with the finite spatial resolution in
computation via MPB. It is found that at small spatial resolution the
Dirac points are splitted. The splitting is less than
5\% of $\ome_D$ for resolution greater than 16. For resolution 24, the
splitting is below 1\% of $\ome_D$. These results indicate that fabrication
error within 5\% of the lattice constant is well sufficient to preserve the
Dirac points, which can be well achieved in the state-of-art fabrication
methods of photonic crystals.\cite{revphc}}

\section{Conclusion and perspectives}
We proposed the
first realization of topological $Z_2$ DPs as 3D photonic analog of graphene
in all-dielectric PCs. We showed that they can exist
only in hexagonal PCs with $C_6$ symmetry. Future research efforts
need to be devoted to the novel properties of 3D photonic
graphene as well as its applications, though some of them are
discussed in this work. At present stage several applications can be
conceived: First, it can be used as 
effective zero-refractive-index medium without loss.\cite{zim} Second,
it can exploited for frequency-, angle-, wavevector- and AM-
selective transmission.\cite{yao} Third, the Imbert-Federov effect
depends on the winding geometry of AM around the WPs as shown in
Ref.~\onlinecite{hua2}. The Imbert-Federov effect will be qualitatively
distinct for different WPs in our PCs. Fourth, the DPs can be used
to design other photonic topological states, such as 3D topological
insulators of light. Fifth, the DPs can be used as a mechanism to
suppress back-scattering and enhance signal transmission length in
communications.\cite{hua} This mechanism will be especially valuable
if a PC fiber design with DPs is invented. 

{We also remark that if an electronic version of our
model is realized, there will be a pair of stable $Z_2$ {\em double}
Dirac points, each of which has {\em eight-fold} degeneracy (including
two-fold spin degeneracy). These novel materials are discussed only very
recently.\cite{kane2} Finally, we emphasize that DPs realize all those
applications without breaking TRS and IS, which considerably reduces
material and fabrication difficulties.}

{\it Note Added:} {Shortly after finalization of this work, a proposal
exploiting a pair of DPs on the $K$-$H$ line of 
a hexagonal PC with $C_6$ symmetry as the mother state toward 3D
all-dielectric weak topological insulator of light was raised.\cite{kiv}}

\section*{Acknowledgments}
 We thank supports from the faculty 
start-up funding of Soochow University and the National Science
Foundation of China for Excellent Young Scientists (grant
no. 61322504). J.H.J thanks Sajeev John, Suichi Murakami and Xiao Hu
for helpful discussions.

\appendix

\section{The ${\vec k}\cdot{\vec P}$ theory for photonic energy bands}

We use the ${\vec k}\cdot{\vec P}$ theory on the study of the photonic
band structure near the $\Gamma$-$A$ line (i.e., we consider
$k_\pp\equiv |{\vec k}_\pp|$ with ${\vec k}_\pp=(k_x,k_y)$ to be
small). The eigenvalue problem in photonic band structure is to solve
the following Maxwell's equations 
\be
\grad\times \frac{1}{\vep({\vec r})} \grad\times {\vec
  H}_{n,\vec{k}}({\vec r}) = \frac{\ome_{n,\vec{k}}^2}{c^2} {\vec
  H}_{n,\vec{k}}({\vec r}) ,
\ee
where $n$ is the band index and ${\vec H}_{n,\vec{k}}({\vec r})$ is
the Bloch function of the magnetic field of photon. The Bloch function
is normalized as $\int_{u.c.} d{\vec r} {\vec
  H}_{n^\prime,\vec{k}}^\ast({\vec r}){\vec H}_{n,\vec{k}}({\vec
  r})=\delta_{nn^\prime}$ with $u.c.$ denoting the unit cell (i.e.,
integration in a unit cell). The Hermitian operator $\grad\times
\frac{1}{\vep({\vec r})} \grad\times$ can be viewed as the photonic
Hamiltonian. 

The essential idea of the ${\vec k}\cdot{\vec P}$ theory around the
$\Gamma$-$A$ line is to expand the Bloch wave function at ${\vec k}$
with the wavefunction at $(0,0,k_z)$ which is denoted as ${\vec
  H}_{n,0,k_z}({\vec r})$. Using such expansion, we obtain the
following ${\vec k}\cdot{\vec P}$ Hamiltonian,
\begin{widetext}
\be
{\cal H}_{nn^\prime} ({\vec k}) = \frac{\ome_{n,0,k_z}^2}{c^2}\delta_{nn^\prime} +
{\vec k}\cdot {\vec P}_{nn^\prime}  - \int_{u.c.} 
\frac{d{\vec r}}{\vep({\vec r})} \vec{H}_{n,0,k_z}^\ast({\vec
  r})\cdot[\vec{k}\times(\vec{k}\times \vec{H}_{n^\prime,0,k_z}({\vec r}))] ,
\ee
where $\ome_{n,0,k_z}$ is the eigen-frequency of the $n^{th}$ band at the
$\Gamma^\prime=(0,0,k_z)$ point. The matrix element of ${\vec P}$ is given by 
\be
{\vec P}_{nn^\prime} = \int_{u.c.}\frac{d\vec{r}}{\vep({\vec r})} [\vec{H}_{n^\prime,0,k_z}({\vec
  r})\times(i\grad\times \vec{H}_{n,0,k_z}^\ast({\vec r}))+(i\grad\times
\vec{H}_{n^\prime,0,k_z}({\vec r}))\times \vec{H}_{n,0,k_z}^\ast({\vec
  r})]  .
\ee
\end{widetext}
We notice that the matrix element of ${\vec P}$ is nonzero only when
the $n$ and $n^\prime$ bands are of different parity. Moreover, the
(approximate) conservation of orbital angular momentum along $z$
direction $L_z$ dictates that the ${\vec P}$ matrix element is
prominent only between bands of angular momentum difference $\pm
1$. Therefore, the coupling between the $p$ and $d$ bands are within
the same pseudo spin, that is, between $p_+$ and $d_+$ ($L_z=\hbar,2\hbar$,
respectively), or between $p_-$ and $d_-$ ($L_z=-\hbar,-2\hbar$).

\begin{figure}[htb]
  \begin{center}
    \includegraphics[width=8.cm]{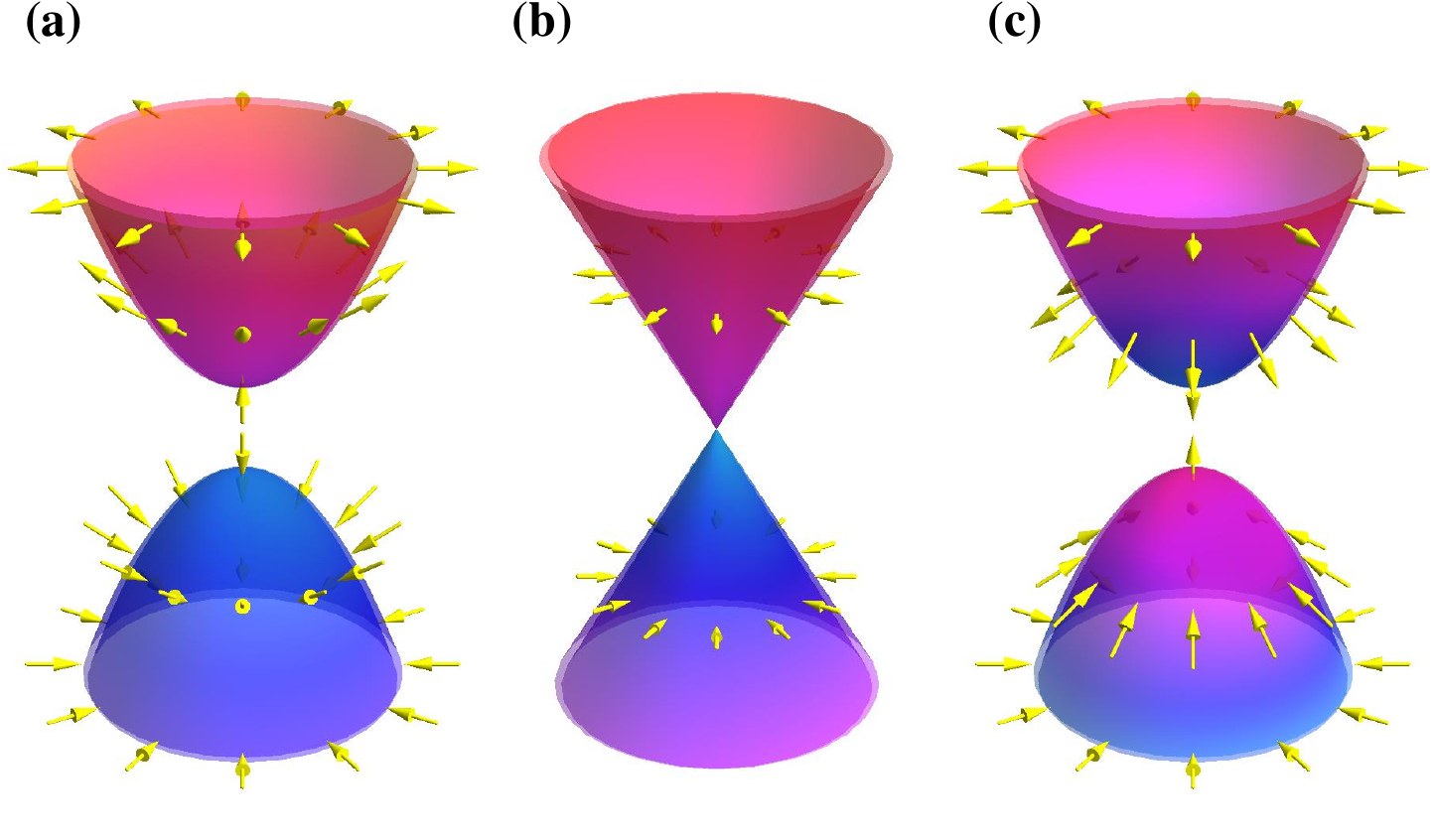}
    \caption{ Photonic dispersion in $k_x$-$k_y$ plane around the
      $(0,0,k_z)$ point for (a) $k_z<K_z$, (b) $k_z=K_z$, and (c)
      $k_z>K_z$. There are two nearly degenerate bands above and below the
      Dirac point. The arrows indicate the orbit angular momentum
      distribution for the pseudo-spin up bands. Note that the orbit
      angular momentum has been substracted by an universal constant
      $\frac{3}{2}\hbar$ (i.e., the average orbit angular momentum
      between the $p_+$ and $d_+$) state.}\label{fig:S2}
  \end{center}
\end{figure}

Using the symmetry properties of the $|p_x\rangle$, $|p_y\rangle$, $|d_{x^2-y^2}\rangle$ and
$|d_{xy}\rangle$ wavefunctions and the $P_x$ and $P_y$ operators, one can find
that\cite{sakoda,hu} $\langle p_x|P_x|d_{x^2-y^2}\rangle=\langle
p_x|P_y|d_{xy}\rangle=\langle p_y|P_x|d_{xy}\rangle=-\langle
p_y|P_y|d_{x^2-y^2}\rangle$. Therefore, $\langle p_+|{\cal
  H}|d_+\rangle=\langle p_-|{\cal H}|d_-\rangle^\ast = A_\pp k_\pp
e^{i\theta_k}$, where $\theta_k\equiv {\rm Arg}[k_x+ik_y]$ and $A_\pp
\equiv\langle p_x|P_x|d_{x^2-y^2}\rangle$. Thus the ${\vec k}\cdot{\vec P}$
Hamiltonian written in the basis of $(d_+,p_+,d_-,p_-)^T$ is
\be
{\cal H} = \frac{2\ome_0}{c^2}\left( \begin{array}{cccccccccccc}
    \frac{\ome_d^2({\vec k})}{2\ome_0} & v_{\parallel} k_{\parallel} e^{-i\theta_k} & 0 & 0  \\
    v_{\parallel}^\ast k_{\parallel}e^{i\theta_k} & \frac{\ome_p^2({\vec k})}{2\ome_0} & 0 & 0 \\
      0 & 0 & \frac{\ome_d^2({\vec k})}{2\ome_0} & v_{\parallel}^\ast k_{\parallel} e^{i\theta_k} \\
      0 & 0 & v_{\parallel} k_{\parallel}e^{-i\theta_k} & \frac{\ome_p^2({\vec k})}{2\ome_0} \\
    \end{array}\right) ,
\ee
where $\ome_0$ is the frequency of the Dirac point and 
$v_\pp\equiv \frac{A_\pp^\ast c^2}{2\ome_0}$. The group velocity in
the $x$-$y$ plane is $|v_\pp|$ which is a function of $k_z$. To the
lowest nontrivial order in ${\vec k}_\pp$ and $k_z$, $\ome_d({\vec k})=\ome_{d,0}(k_z)$ and
$\ome_p({\vec k})=\ome_{p,0}(k_z)$. 

The condition $\ome_{p,0}(k_z)=\ome_{d,0}(k_z)$ is satisfied only at
$k_z=\pm K_z$ (more rigorously, the two points $(0,0,\pm K_z)$ are the
only points where the $p$ and $d$ bands become degenerate).
The topological phase transistion as a function of $k_z$ is
illustrated in Fig.~\ref{fig:S2}. For $k_z<K_z$ the band structure and spin
configuration resembles that of a Dirac electron with negative mass.
It is known that the negative mass Dirac equation describe the $Z_2$
topological insulator in electronic system.\cite{rev1,rev2}
For $k_z=K_z$ the band gap closes and a Dirac cone emerges. For
$k_z>K_z$ the cone is gaped again where the spin configuration
resembles that of a Dirac electron with positive mass.

When the six micropillars are twisted, the symmetry of the photonic
crystal is reduced from $C_{6v}\otimes I_z$ to $C_6$ ($I_z$
is the inversion along the $z$ direction). Since $I_z$ is broken, the
spectrum for $k_z>0$ is no longer the mirror of that for $k_z<0$.
However, the TRS guarantees the degeneracy between the
$|p_+,k_z\rangle$ state and the $|p_-,-k_z\rangle$ state as well as
the degeneracy between the $|d_+,k_z\rangle$ state and the
$|d_-,-k_z\rangle$ state. Since the $C_6$ symmetry is kept, those four
bands can cross each other at ${\vec k}_\pp=0$, as they correspond to
different eigenvalues of the $C_6$ operator. Therefore, along the
$\Gamma$-$A$ line the following ``mass terms'' are introduced
$\sim m_1 \hat{\sigma}_z + m_2 \hat{\tau}_z + m_3
\hat{\tau}_z\hat{\sigma}_z$ (the three quantities, $m_1$,
$m_2$, and $m_3$ are the ``masses'') which gap out
the Dirac point. Those mass constant are odd functions of $k_z$ due to
TRS. As a consequence, the DPs are split into WPs with chirality shown
in Fig.~3.

\section{Mie resonances in hollow cylinder and their connection with photonic energy bands}

\begin{figure}[htb]
  \begin{center}
    \includegraphics[width=8.5cm]{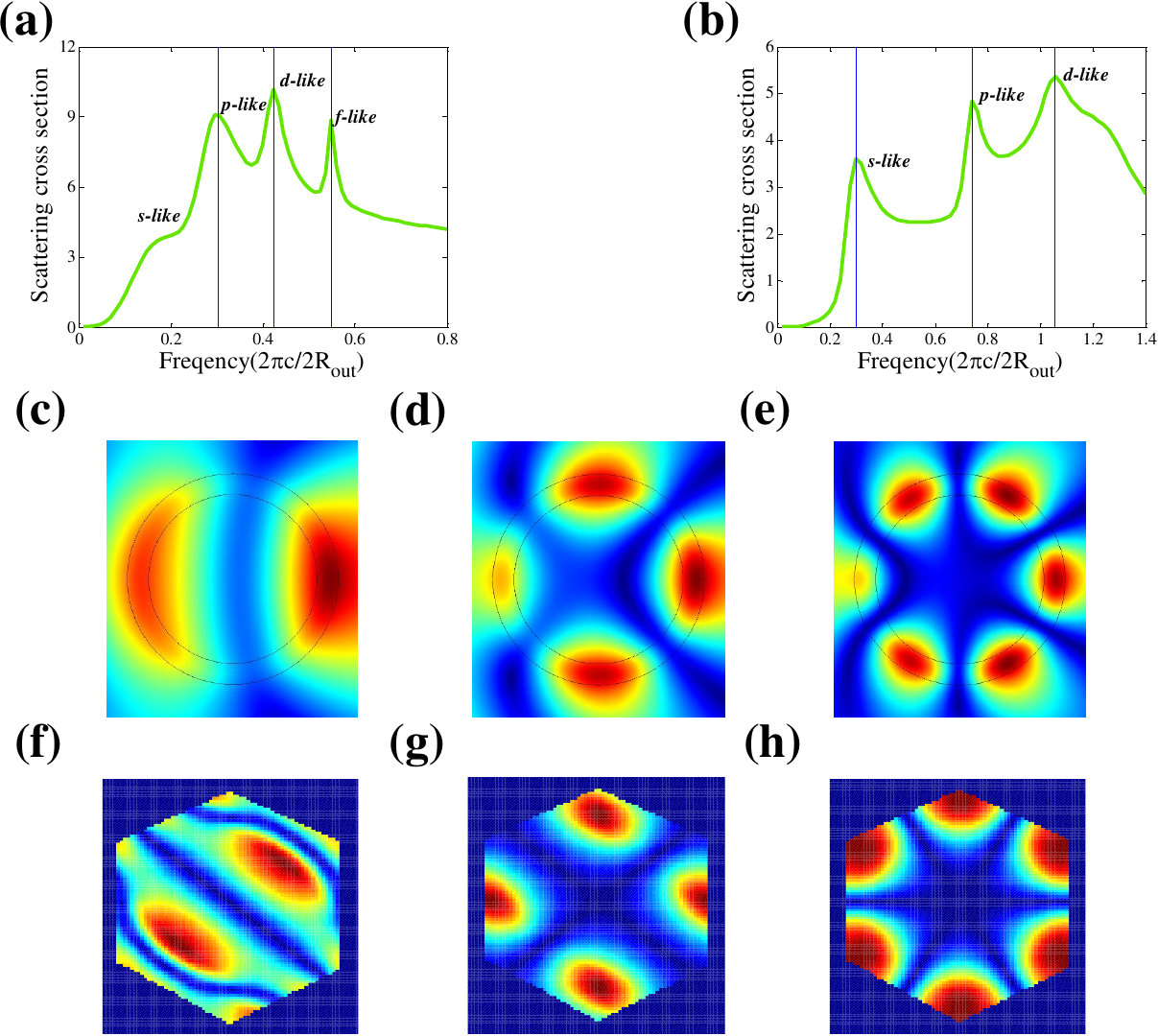}
    \caption{ Mie resonances of infinitely long hollow cylinder. a, The
      scattering cross section as a function of the frequency of the
      incident plane wave with TM polarization ($E_z$ polarization,
      $z$ direction is along the cylinder). b, The scattering cross
      section vs. the frequency of the incident plane wave with TE
      polarization ($H_z$ polarization). c,d,e, The electric field
      patterns $|E_z|$ at the first, second, and third
      resonances for the TM polarization, respectively. They are
      identified as the $p$-, $d$-, and $f$-wave scatterings. f.g.h.,
      The electric field $|E_z|$ patterns in the $x$-$y$ plane in a unit cell (averaged
      over the $z$ direction) for the $p$, $d$, and $f$ states in the
      hollow cylinder photonic crystal, respectively.}\label{fig:S1}
  \end{center}
\end{figure}

When light is scattered by dielectric object, there are geometry
induced resonances which appear in the frequency dependence of the
scattering cross-section. This is known as the Mie resonance. The
rigorous connection between the Mie resonance and the photonic 
bands was systematically established by Lidorikis et al.\cite{soukoulis}
Inspired by the fact that the Mie resonance
frequencies are related to the photonic energy bands, those authors
extended the idea of the linear combination of atomic orbitals method
to photonic energy bands. In their theory the Mie resonances of
isolated cylinder are treated as the atomic orbitals of photonic
crystals. The hybridization/transfer between neighboring localized
resonances leads to the formation of the photonic band. Particularly,
in a tight-binding theory, one can expand the photonic wavefunction as
superposition's of the plane wave and the Mie resonances, 
\be
\vec{\Psi}_{n,{\vec k}} = \frac{\alpha_n}{\sqrt{V}} e^{i{\vec k}\cdot{\vec r}} + \frac{\beta_n}{\sqrt{V}}
\sum_{\vec{R}}\vec{\varPsi}_n({\vec r}-{\vec R})e^{i{\vec k}\cdot{\vec R}}
\ee
where $\vec{\varPsi}_n({\vec r}-{\vec R})$ stands for the wavefunction
of the Mie resonance at the ${\vec R}$ lattice site (${\vec R}$ is the
position of the center of that lattice site), $\alpha_n$ and
$\beta_n$ denote the coefficient for the plane wave and the Mie
resonance, respectively. The index $n$ here labels the Mie resonance,
from $s$ to $p_x$, $p_y$, etc. $V$ is the volume of the photonic
crystal. The plane wave component is nonzero
only for the $n=0$ (i.e., the $s$-wave case). The plane wave component
is a unique feature of the photonic bands, because it dominates the
photonic wavefunction in the long wave length (i.e., low frequency)
limit. It has been shown by Lidorikis et al.\cite{soukoulis} that the above
wavefunction can be used as basis to expand the wavefunctions of the
photonic energy bands. For example, the first 
photonic bands is plane-wave-like for small $|{\vec k}|$, while it
gradually becomes $s$-wave-like and mix with other resonances at large
$|{\vec k}|$. Higher photonic bands are dominated by the Mie
resonances. A solid connection between Mie scattering
cross-section, resonance frequency and the scaling function of the
tight-binding parameters is also established in the work of Lidorikis
et al.\cite{soukoulis}

In Fig.~\ref{fig:S1}a the frequency dependence of scattering cross section is shown for a hollow
cylinder with outer radius 0.5$a$ and inner radius 0.4$a$ (the same as the
hollow cylinders in our photonic crystal) for the TM
polarization (calculated via COMSOL MultiPhysics). Fig.~\ref{fig:S1}b gives the results for the TE polarization. In
both calculation the cylinder is assumed to be infinitely long, and
the photon wave vector is in the $x$-$y$ plane. Thus, the problem
considered is a two-dimensional problem.

The scattering cross-section as a function of the frequency of the
incident plane wave is shown in Figs.~\ref{fig:S1}a and 1b for the two
situations when the incident wave is TM polarized (i.e., $E_z$
polarized) and when it is TE polarized (i.e., $H_z$ polarized),
respectively. There are several peaks in the two figures which are the
recognized as the Mie resonances. The electric field $|E_z|$ patterns
at the resonant frequency's are shown for the TM polarization in
Figs.~\ref{fig:S1}c, 1d, and 1e. The field patterns show $p, d$, and
$f$-wave symmetry for the second, third, and fourth resonances for the
TM polarization. We notice that the TE resonances have much higher
frequency than the TM resonances. The field patterns of the TM Mie
resonance is comparable with that of the $p$, $d$ and $f$ photonic
states in the hollow cylinder photonic crystal (see
Figs.~\ref{fig:S1}f, 1g, and 1h). The resonant frequency's for the
$p$- and $d$-wave scatterings for the TM polarization are $0.3$ and
$0.42$, respectively. These numbers, multiplied by the square root of
the effective permittivity in our three-dimensional photonic crystal,
$\vep_{eff}=f\vep_{silicon}+(1-f)\vep_{air}=1.9$, with $f=0.236$
($\vep_{silicon}=12$ is the permittivity for the dielectrics, e.g.,
silicon) being the filling ratio of silicon,\cite{soukoulis} yield
0.57 and 0.8 $\frac{2\pi c}{a}$, respectively. This rough estimation
gives quite comparable results with the frequency's of the $p$- and
$d$- photonic bands at the $A$ point of our photonic crystal, $0.53$
and $0.65 \frac{2\pi c}{a}$, respectively.

\section{Dirac points without micropillars}

\begin{figure}[htb]
  \begin{center}
    \includegraphics[width=8.6cm]{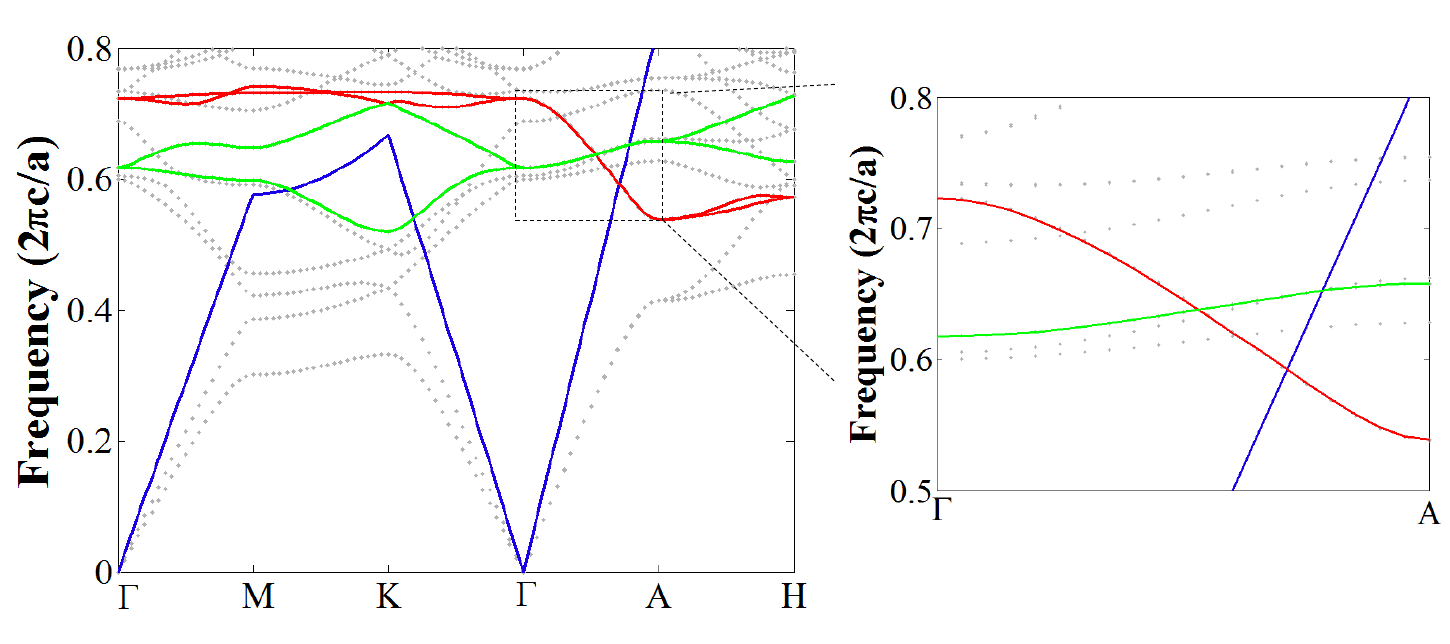}
    \caption{ Photonic band structure of hexagonal photonic crystal
      without micropillars. The height of the unit cell is 0.6$a$. The
      green curve represents the $p$ bands, whereas the red curve for
      the $d$ bands. The blue curve denotes the light-line. The
      height of the hollow cylinder is 0.4$a$ (there is no micropillars). Other parameters: $R_{out}/a=0.5$
      and $R_{in}/a=0.4$, and $\vep=12$.}\label{fig:S11}
  \end{center}
\end{figure}

Here we show that the Dirac points exist even when the six
micropillars are removed. The simplified structure is maybe more
attractive for applications. The photonic bands are shown in
Fig.~\ref{fig:S11}. It is seen that the Dirac points also emerges as an
accidental crossing of the doubly-degenerate $p$ and $d$ bands above
the light-line. Meanwhile there is a Dirac point emerge on the
$K$-$H$ and $K^\prime$-$H^\prime$ line. Although there is only
two-fold degeneracy at the $K$ point from the figure, taking into
account of the degeneracy at the $K^\prime$ point with the same
frequency, in total it is four-fold degeneracy, i.e., a Dirac point.

\section{Symmetry analysis}
We now analyze the constraints of the time-reversal symmetry (TRS),
inversion symmetry (IS) and $C_6$ rotation symmetry on the photonic
Hamiltonian. Let us look at the following form of a $4\times 4$
Hamiltonian,
\be
H({\vec k}) = \left( \begin{array}{cccc} h_{\up\up} ({\vec k}) &
    h_{\up\down}({\vec k}) \\
    h_{\down\up} ({\vec k}) &
    h_{\down\down}({\vec k}) \end{array} \right),
\ee
where each $h$ is a $2\times 2$ matrix which is a linear combination
of $\tau_0$, $\tau_x$, $\tau_y$ and $\tau_z$ ($\tau_0$ is identity matrix). 

First, the pseudo-time-reversal symmetry $H(-{\vec k})=\Theta H({\vec
  k})\Theta^{-1}$ with $\Theta=i\sigma_yK$ gives rise to the following
constraints, $h_{\down\down}({\vec k})=-h_{\up\up}^\ast(-{\vec k})$ and
$h_{\up\down}({\vec k})=h_{\up\down}^T(-{\vec k})$ where $T$ denotes
transposition. We can thus describe the total Hamiltonian using only
$h_{\up\up}$ and $h_{\up\down}$. The pseudo-time-reversal
symmetry can always be defined whenever there is a two-fold
degeneracy.\cite{hou} Ref.~\onlinecite{hu} showed how to construct the
pseudo-time-reversal symmetry for the $p$ and $d$ bands in $C_6$
symmetric photonic crystals.

Second, the inversion symmetry is $H(-{\vec k})={\cal P} H({\vec k})
{\cal P}^{-1}$ with ${\cal P}=\tau_z$. This imposes the further
restrictions, $h_{\up\up}({\vec k})=\tau_z h_{\up\up}(-{\vec
  k})\tau_z$ and $h_{\up\down}({\vec k})=\tau_z h_{\up\down}(-{\vec
  k})\tau_z$. 

We remind the readers that in construction the pseudo-time-reversal
operator and the inversion operator we have already used the
properties that the four bands are the eigen-representation of the
$C_6$ symmetry with opposite parity. Using the ${\vec k}\cdot {\vec
  P}$ theory up to linear terms in ${\vec k}$, one finds that the
coupling between different bands is only within each pseudo-spin.
This coupling between bands with opposite parity has a winding phase
in $k_x$-$k_y$ plane in order to conserve the angular momentum along
$z$ direction.\cite{fang,nagaosa} Hence, up to linear terms in ${\vec
  k}$,
\begin{align}
& h_{\up\up}({\vec k}) = c_1(k_z) \tau_0 + c_2(k_z) \tau_z \nn\\ & \hspace{1.8cm} + [ c_3(k_z)
(k_x \pm i k_y)\tau_+ + {\rm H.c.} ] ,  \\
& h_{\up\down}({\vec k}) = 0 ,
\end{align}
where $c_1(k_z)$, $c_2(k_z)$, and $c_3(k_z)$ are even functions of
$k_z$. The sign $\pm$ depends on the definition of the pseudo-spin.  

According to Ref.~\onlinecite{nagaosa}, the $C_6$ symmetry can also stabilize
Dirac points the off-diagonal terms in $h_{\up\up}({\vec k})$ is of
the form $\sim k_z (k_x\pm ik_y)^2$. Such coupling is impossible for
photonic bands, because it is of the form between bands with same
parity in $x$-$y$ plane which will anti-cross at a generic ${\vec k}$
point along the $\Gamma$-$A$ line (since at such ${\vec k}$ point
there is only $C_6$ symmetry, the two doubly-degenerate bands with
same parity will belong to the same representation, hence $C_6$
symmetry cannot prevent them from anticrossing).

Therefore, in photonic crystals the only possible Dirac points is of
the same form as what we discussed in the main text. The other
possible case is for the Dirac points on the $K$-$H$ line. The
simplest way to understand this is that by enlarging the unit cell in
real space, the $K$-$H$ line folds back to the $\Gamma$-$A$
line.\cite{hu} Therefore, the situations are similar to what was
discussed in the main text.

\begin{widetext}
\section{Topological surface states}

\begin{figure}[htb]
  \begin{center}
    \includegraphics[width=14.cm]{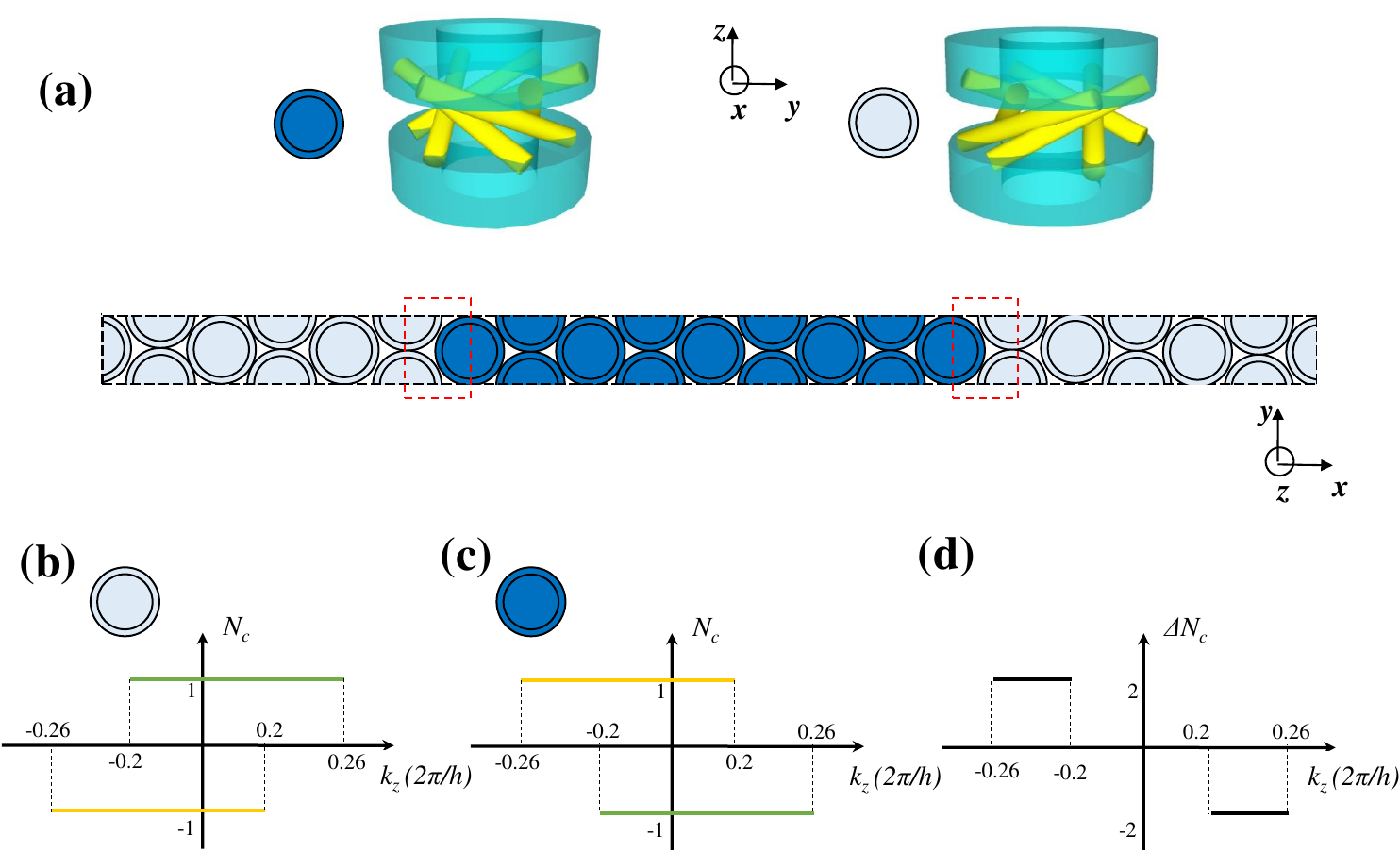}
    \caption{ a, Supercell structure for the calculation of the edge
      states. In the middle is the photonic crystal with micropillars
      twisted clockwise (represented by  blue circles). In the
      other regions are the photonic crystal with anticlockwise
      twisted micropillars (depicted as gray circles). The structure
      of the two types of photonic crystals are shown above the
      supercell. For a given supercell we calculate the photonic band
      structure with a given $k_z$ for various $k_y$ ($k_x=0$ as the
      system has finite size along the $x$ direction). b, Chern
      numbers for the $p_+$ and $d_+$ bands (green) and for the $p_-$ and $d_-$
      bands (yellow) below the Weyl points as functions of $k_z$ for the
      photonic crystal with micropillars twisted clockwise. c, Chern numbers vs. $k_z$ for
      the photonic crystal with anticlockwise twisted
      micropillars. d, Total Chern number difference $\Delta N_C=N_C({\rm
        clockwise})-N_C({\rm anticlockwise})$ as a function of $k_z$.}\label{fig:S6}
  \end{center}
\end{figure}

\begin{figure}[htb]
  \begin{center}
    \includegraphics[width=14.cm]{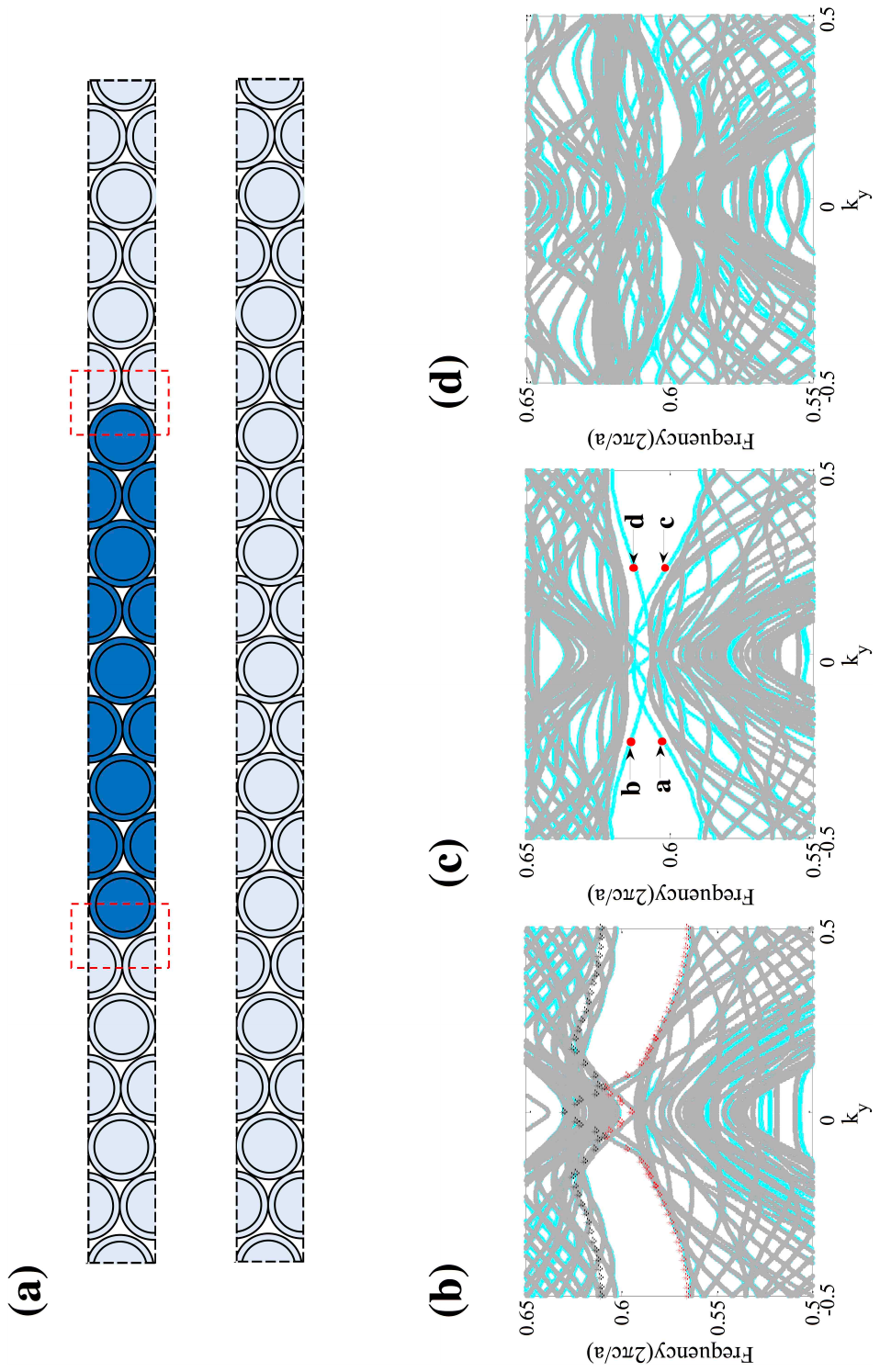}
    \caption{ (a), The supercell structures calculated. Up: the
      heterostructure with two types (clockwise and anticlockwise
      twisted micropillars) of photonic crystals. Down: the structure
      with {\em only one} type of photonic crystal (the one with
      anticlockwise twisted micropillars). (b), The photonic spectrum
      with $k_z=0$ for various $k_y$. The blue curves give the
      spectrum for the heterostructure, while the gray curves give the
      spectrum for the supercell with only one type of photonic
      crystal [the same color scheme is applied to (c) and (d), as
      well]. The gray curves are on top of the blue curves for (b),
      (c), and (d). The projected bulk photonic bands are depicted by
      the red and black points for the upper and lower band edges,
      respectively. (c) and (d), The photonic spectrum for the
      heterostructure and uniform structure with
      $k_z=0.225\frac{2\pi}{h}$ and $k_z 
      =0.3\frac{2\pi}{h}$, respectively. There are four edge states in
      (c). We label four special points a, b, c, and d, with $k_y=\pm
      0.2\frac{2\pi}{a}$. The properties of these four points will be
      shown in Fig.~\ref{fig:S8}.}\label{fig:S7}
  \end{center}
\end{figure}
\end{widetext}

Due to the existence of the $f$-bands, there is no complete band gap
for the hollow cylinder photonic crystal with inversion
symmetry. Nevertheless, for the inversion symmetry broken photonic
crystal, there is complete photonic band gap for
$k_z\lesssim0.3\frac{2\pi}{h}$. We calculate the edge states for the
inversion symmetry broken photonic crystal using the supercell
illustrated in Fig.~\ref{fig:S6}(a). Two armchair boundaries are formed
between two types of photonic crystals with clockwise and
anticlockwise twisted micropillars. Edge states emerge at the two
boundaries as governed by the band topology. According to the ${\vec
  k}\cdot{\vec P}$ analysis, the Chern numbers for the $p_+$ and $d_+$
and for the $p_-$ and $d_-$ bands below the Weyl points are plotted in
Fig.~\ref{fig:S6} for the two 
twisted-micropillar structures. For the clockwise-twisting structure,
the Chern numbers are shown in Fig.~\ref{fig:S6}(b). For the
anticlockwise-twisting structure, the dispersion of the $p_+$-$d_+$
bands are switched with the $p_-$-$d_-$ bands. Thus, for both the
$k_z>0$ and $k_z<0$ regions, the position of 
the Weyl point with positive chirality switches with the position of the
Weyl point with negative chirality. This modifies the $k_z$-dependence
of the Chern number [see Fig.~\ref{fig:S6}(c)]. The difference of the total Chern
number between the two types of photonic crystals,
$\Delta N_C=N_{C}({\rm clockwise})-N_C({\rm anticlockwise})$, is
plotted in Fig.~\ref{fig:S6}(d) as a function of $k_z$. The two types of photonic
crystals have perfectly matched photonic band gap which is suitable
for the calculation of the edge states.

\begin{figure}[htb]
  \begin{center}
    \includegraphics[width=8.5cm]{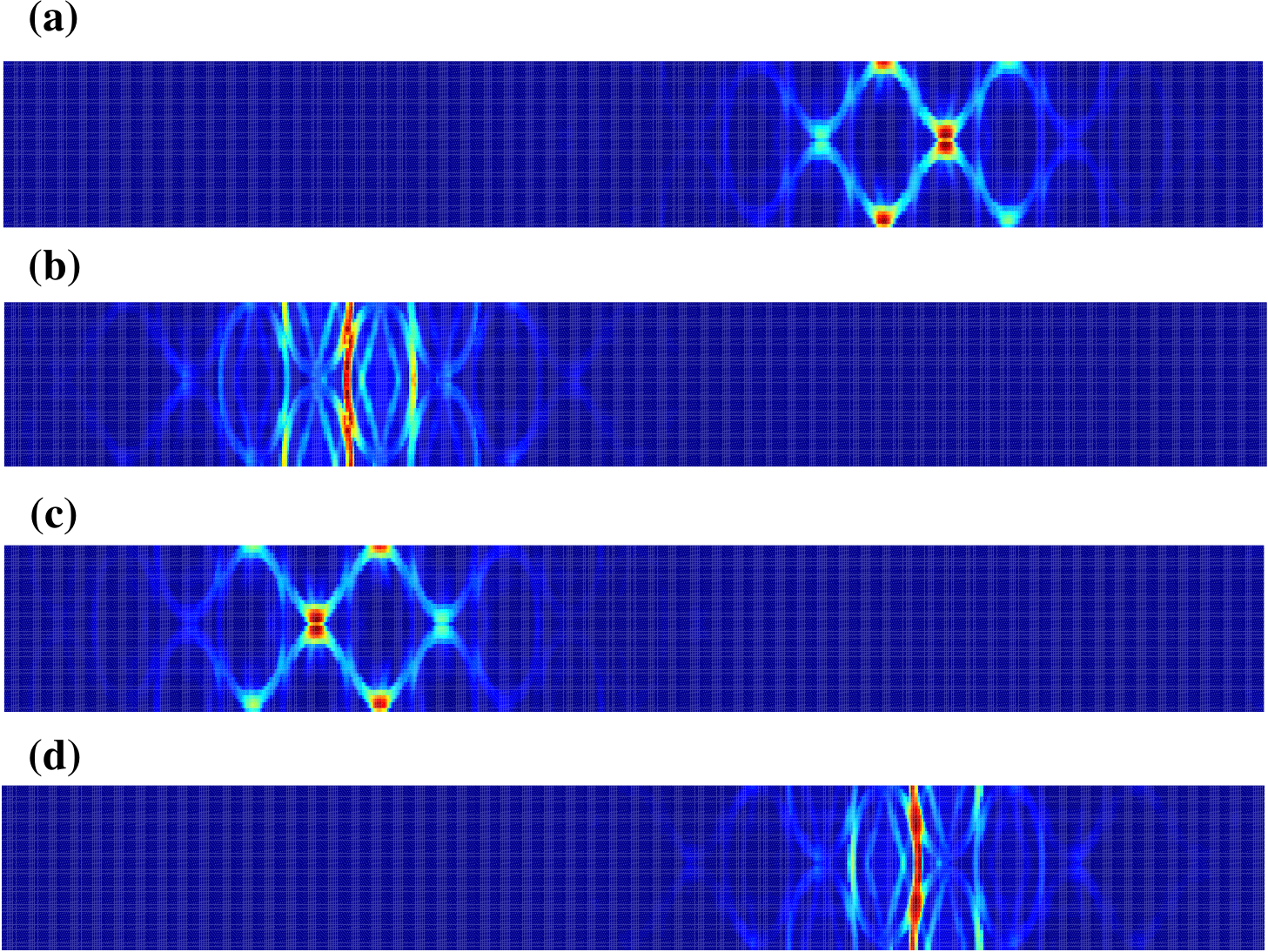}
    \caption{ Electromagnetic energy density $\int dz
      \vep({\vec r})|\vec{E}({\vec r})|^2$ in the $x$-$y$ plane for
      the a, b, c, and d points in the photonic spectrum in
      Fig.~\ref{fig:S7}(c).}\label{fig:S8}
  \end{center}
\end{figure}

We calculate the edge states for $k_z=0, 0.225\frac{2\pi}{h}$, and
$0.3\frac{2\pi}{h}$ for a supercell with 9 periods of
clockwise-twisting structures and 11 periods of
anti-clockwise-twisting structures along the $x$ direction [see
Fig.~\ref{fig:S7}(a)]. According to Fig.~\ref{fig:S6}(c), the Chern number difference is
zero for $k_z=0$ and $0.3\frac{2\pi}{h}$, while $\Delta N_C=-2$ for
$k_z=0.225\frac{2\pi}{h}$. To carefully take into account of the
finite size effect in the supercell calculation (as the band gap can
be very small here), we plot the photonic spectrum for the
heterostructure (blue) together with the spectrum for the supercell
with the same size but with only the anti-clockwise-twisting 
structures (gray) [The two kinds of supercell are illustrated in
Fig.~\ref{fig:S7}(a)]. Since the supercell with only the anti-clockwise-twisting 
photonic crystal has no boundary, its spectrum acts as a good
reference for the bulk photonic spectrum with the finite size effect
included.

In Fig.~\ref{fig:S7}(b) the blue and gray (with gray on top) spectra are almost
identical. Although the finite size effect closes the band gap near $k_y=0$, the results
show that there is no edge states for $k_z=0$. Note that the
dispersion near $k_y=0$ is not due to edge states. We have checked the
projected bulk photonic bands [red and black points in Fig.~\ref{fig:S7}(b) for
the upper and lower band edges, respectively]. It
is seen that the projected photonic band gap can be very small near
the $k_y=0$ region. The photonic band gap actually closed for finite
size supercell with a single type of photonic crystal (as depicted by
the gray curves). Note that the ``curves'' in Figs.~\ref{fig:S7}(b), 7(c) and
7(d) are actually made of (many) points from the band structure
calculation using MPB. We did not connect them artificially. Fig.~\ref{fig:S7}(c)
clearly demonstrates the existence of edge states for
$k_z=0.225\frac{2\pi}{h}$ in the common photonic band gap.
For $k_z=0.3\frac{2\pi}{h}$ [Fig.~\ref{fig:S7}(d)], the
photonic band gap is very small, we did not notice any edge state.

To further prove the existence of edge states and study their
properties for $k_z=0.225\frac{2\pi}{h}$, we plot the electromagnetic
energy density $\int dz \vep({\vec r})|\vec{E}({\vec r})|^2$ in the
$x$-$y$ plane for the a, b, c, and d points in the photonic spectrum
in Fig.~\ref{fig:S7}(c). From Fig.~\ref{fig:S8} it is seen that there are two edge states at
the right-hand-side boundary, the a and d points. From the photonic
spectrum in Fig.~\ref{fig:S7}(c) we find that the group velocity at both the a
and d points are positive [leading to two anticlockwise (from top-down
view) edge modes]. Therefore, there are two one-way edge
states at the boundaries for $k_z=0.225\frac{2\pi}{h}$, which is in
accordance with the Chern number difference given in Fig.~\ref{fig:S6}(d).
This simulation proves our analysis of the topology of the
photonic bands based on symmetry and the ${\vec k}\cdot {\vec
  P}$ theory.

\section{Spatial field patterns and Poynting vector profiles}

\begin{figure}[htb]
\begin{center}
\includegraphics[width=9cm]{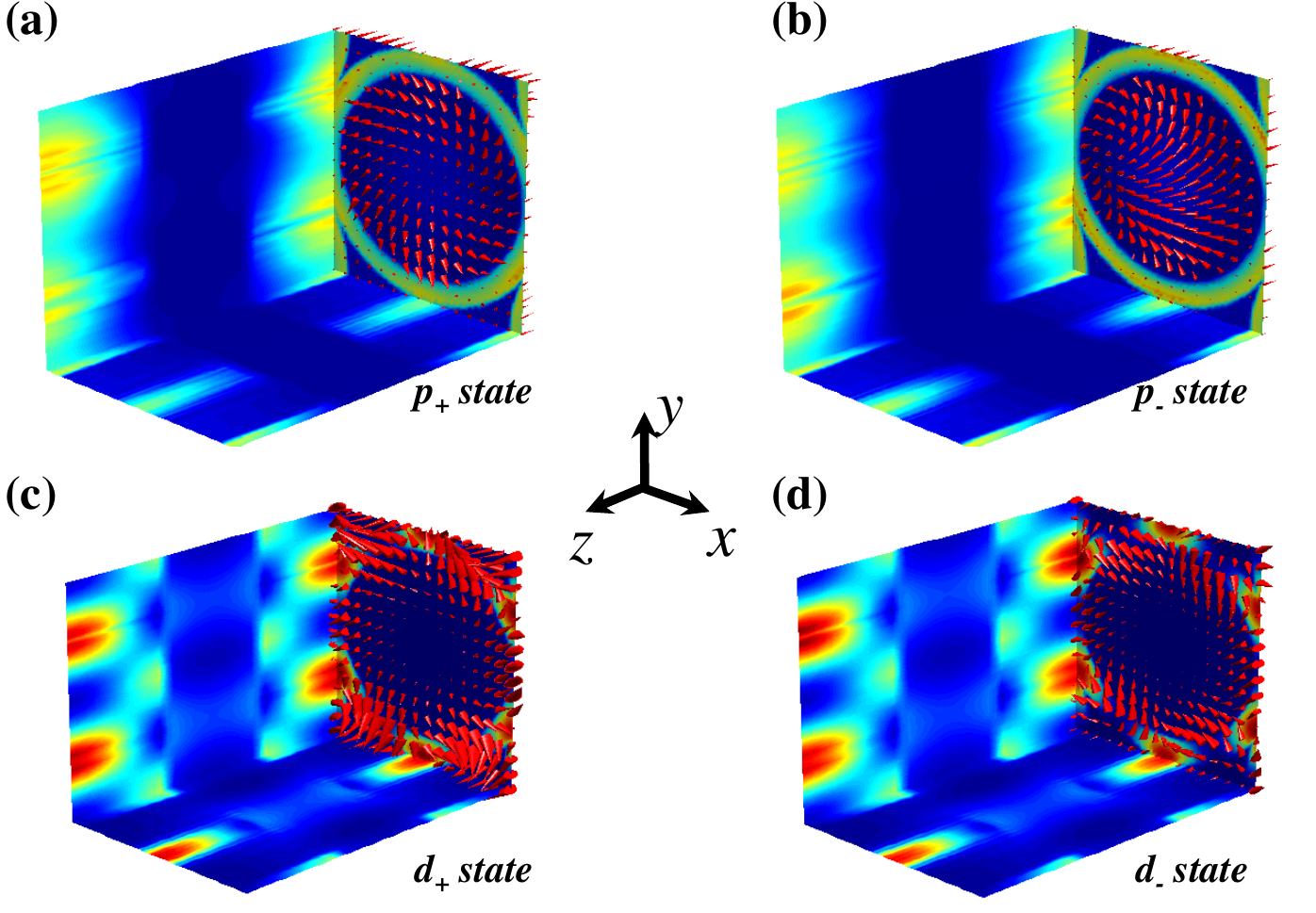}
\caption{ Spatial field patterns on the surfaces of a block that
  contains a unit cell for the $p_{\pm}$ and $d_{\pm}$ bands at
  ${\vec k}=(0,0,0.1\frac{2\pi}{h})$ in chiral photonic crystal. In
  each figure, on the three boundary planes (i.e., the $x$-$y$,
  $y$-$z$, and $x$-$z$ planes) we plot the energy density
  $\vep|\vec{E}|^2({\vec r})$ by color. The red color
  represents high energy density, while the blue denote low energy
  density. The Poynting vector (the time-averaged value $\Re[\vec{E}^\ast\times
  \vec{H}]/2$) distribution is also plotted in the $x$-$y$ plane.}\label{fig:S5}
\end{center}
\end{figure}

In the chiral photonic crystal (shown in Fig.~3)
the four states, $p_+$, $p_-$, $d_+$, and $d_-$, can be identified via
their Poynting vector profile. In Fig.~\ref{fig:S5}, we plot the
spatial profile of the Poynting vector (i.e., the time-averaged value
$\Re[\vec{E}^\ast\times \vec{H}]/2$) as well as that of the
electromagnetic energy density (i.e., $\vep|\vec{E}|^2({\vec r})$) on the
surfaces of a block that contains a unit cell for the four photonic
bands at ${\vec k}=(0,0,0.1\frac{2\pi}{h})$. The spatial patterns of
the Poynting vector are plotted in the $x$-$y$ plane in
Fig.~\ref{fig:S5}. It is noted that the Poynting vector exhibits
spatial distributions similar to skyrmion configuration. In such
skyrmion configuration the Poynting vector winds around the $z$
axis. The winding direction is the same as the pseudo-spin direction,
revealing the finite orbital angular momentum of the four states.
In addition, the Poynting vectors at the center point along $z$
whereas at the boundaries they point along $-z$. Such flip of the
Poynting vector, alongside with the winding texture comprises the full
skyrmion configuration. We remark that those special field profiles
can be used to identify topological photonic bands in chiral photonic
crystals.

In each figure, on the three boundary planes (i.e., the $x$-$y$,
$y$-$z$, and $x$-$z$ planes) we plot the energy density
$\vep|\vec{E}|^2({\vec r})$ by color. The red color
represents high energy density, while the blue denote low energy
density. For example, the energy density distribution in the $x$-$y$
plane has a ring pattern for $p_{\pm}$ states (the ring coincides
with the cross-section of the hollow cylinder). From the figures we
notice that the electromagnetic energy is mostly localized on the
cylinder for both $p$ and $d$ bands. However, the 
energy density profile for the $d$ bands is more anisotropic and
dispersed away from the cylinder.

\section{Moving and Annihilating of $Z_2$ Dirac point}

\begin{figure}[htb]
  \begin{center}
    \includegraphics[width=6.cm]{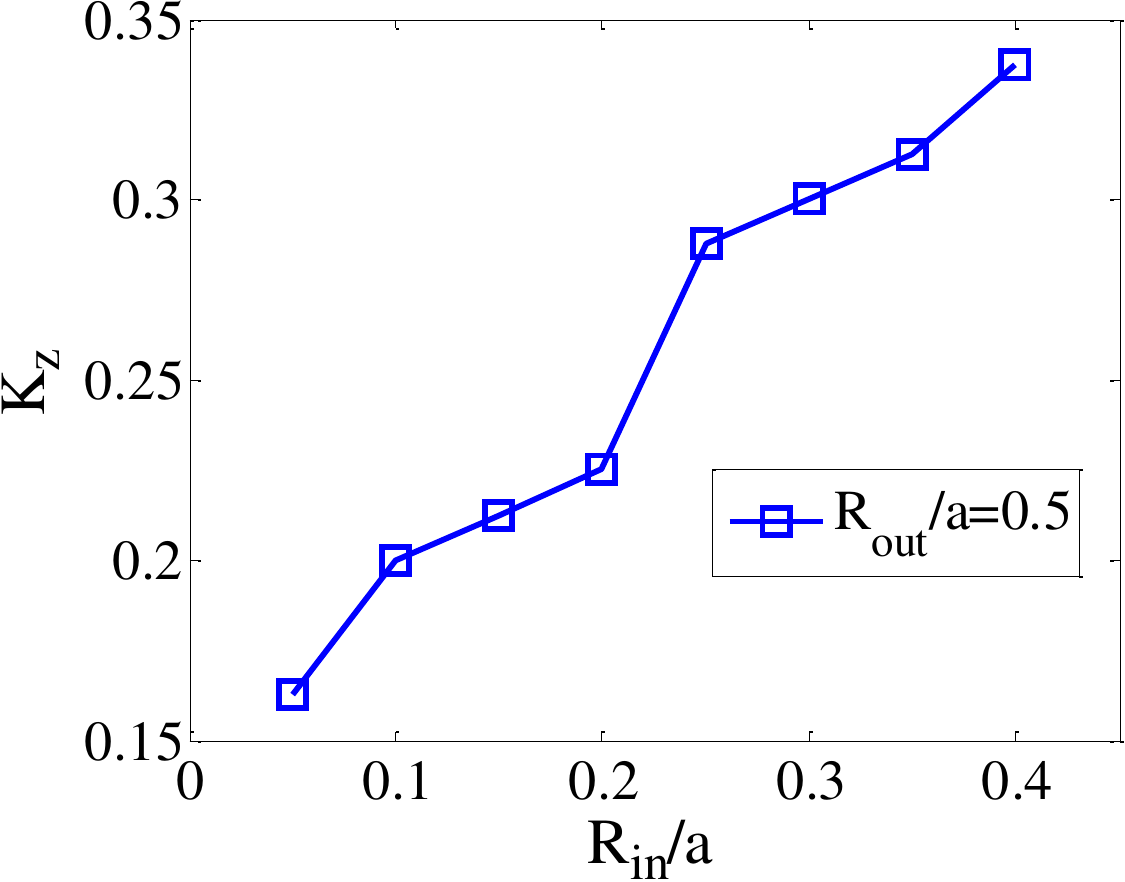}
    \caption{ Dependence of the position of the Dirac point $K_z$
      on the inner radius $R_{in}$ of the hollow cylinder. For small
      $R_{in}$ the Dirac points are moved to the $\Gamma$ point and
      annihilated in pair.}\label{fig:S9}
  \end{center}
\end{figure}

Here we show the moving of the Dirac points when the inner radius of
the hollow cylinder is significantly changed, to demonstrate the
robustness of the Dirac points in our photonic crystal.
In Fig.~\ref{fig:S9}, we plot the position of the Dirac point $K_z$ as
a function of the inner radius $R_{in}$ of the hollow cylinder at
fixed outer raidus $R_{out}$. The two curves represent the case with
$R_{out}=0.5a$. It is clearly seen that $K_z$ decreases with
decreasing $R_{in}$. At very small $R_{in}$ the Dirac points are moved
to the $\Gamma$ point and annihilated there. This confirms that the
$Z_2$ Dirac points can only be created and annihilated in pairs, and
demonstrates the robustness of the $Z_2$ Dirac points.

\begin{figure}[htb]
  \begin{center}
    \includegraphics[width=6.cm]{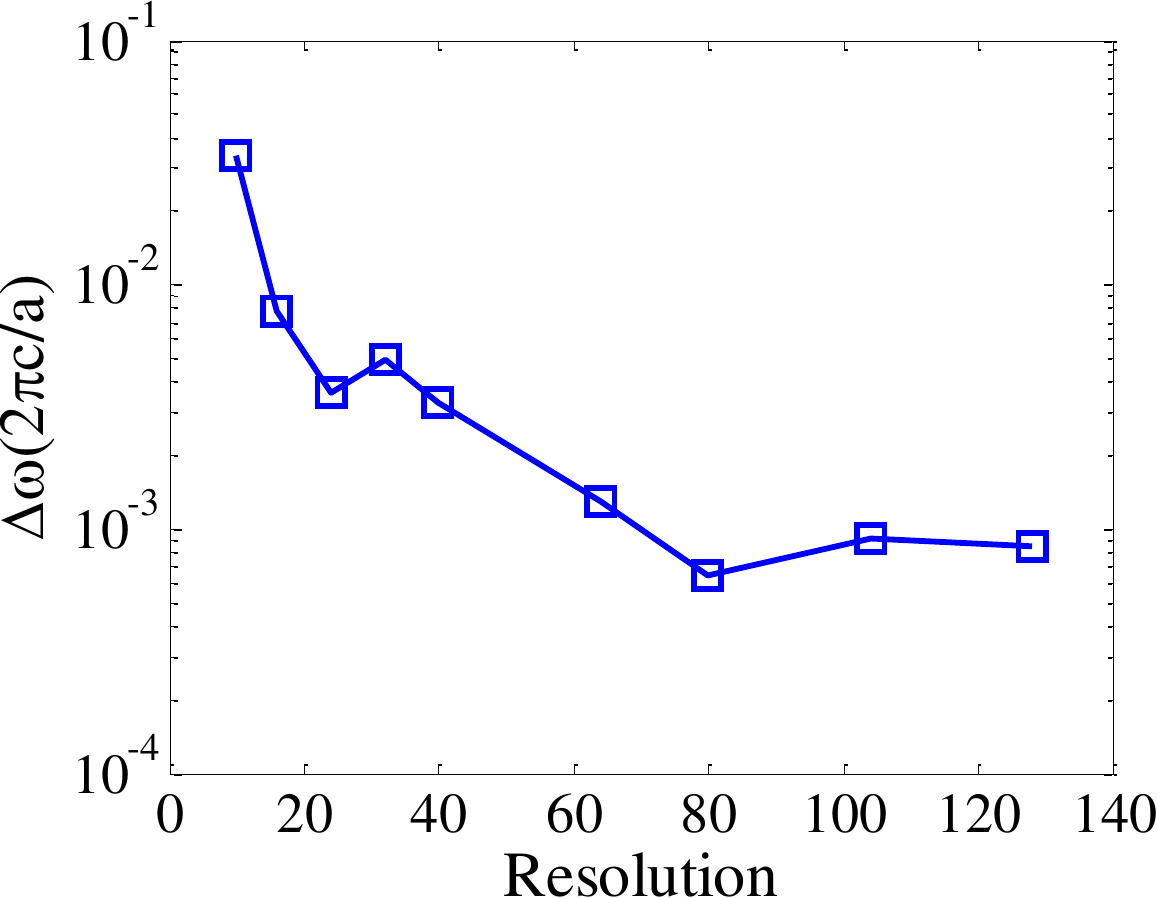}
    \caption{ Splitting of the Dirac points vs. simulation
      resolution. Parameters: $R_{out}/a=0.5$
      and $R_{in}/a=0.4$, and $\vep=12$.}\label{fig:S10}
  \end{center}
\end{figure}

\section{Robustness against weak disorders}

In this section we study the robustness of the Dirac points against
disorder effects such as fabrication errors. We use MPB to calculate
the photonic band structure. In MPB the Maxwell equation is solved via
real space grid method. The space resolution $N$ (i.e., how many grid
points for the distance of the lattice constant $a$) in MPB
calculation is a natural way to simulate the space resolution in
fabrication. We calculate the splitting of the Dirac points (i.e., the
largest frequency minus the smallest one) vs. the
simulation resolution and plot the results in Fig.~\ref{fig:S10}. From the
figure we find that the splitting of the Dirac points is smaller than
5\% of the frequency at the Dirac point $\ome_D$ for resolution
greater than 16. For resolution 24, the splitting is below 1\% of the
frequency at the Dirac point. These results indicate that fabrication
error within 5\% of the lattice constant is sufficient to preserve the
Dirac points, which can be well achieved in the state-of-art fabrication
methods of photonic crystals.\cite{revphc}

{}

\end{document}